\documentclass[useAMS,usenatbib]{mn2e}

\usepackage{graphicx}
\usepackage{amssymb}
\usepackage{multirow}
\usepackage{pifont}

\makeatletter
    
    \newcommand{\Rmnum}[1]{\expandafter\@slowromancap\romannumeral #1@}
\makeatother

\title[A study of the high-inclination population in the Kuiper belt -- \Rmnum3. The 4:7 mean motion resonance]
         {A study of the high-inclination population in the Kuiper belt -- \Rmnum3. The 4:7 mean motion resonance}
\author[Jian Li, S. M. Lawler, Li-Yong Zhou and Yi-Sui Sun]
{Jian Li$^1$\thanks{E-mail: ljian@nju.edu.cn}, S. M. Lawler$^2$,  Li-Yong Zhou$^1$ and Yi-Sui Sun$^1$\\
$^1$School of Astronomy and Space Science
\& Key Laboratory of Modern Astronomy and Astrophysics in Ministry of Education,\\
 Nanjing University, Nanjing 210093, PR China\\
$^2$Campion College, University of Regina, Regina, SK S4S 0A2, Canada}
\begin{document}

\date{Accepted 1988 December 15. Received 1988 December 14; in original form 1988 October 11}

\pagerange{\pageref{firstpage}--\pageref{lastpage}} \pubyear{2002}

\maketitle

\label{firstpage}

\begin{abstract}

The high-inclination population in the 4:7 mean motion resonance (MMR) with Neptune has also substantial eccentricities ($e\gtrsim0.1$), with more inclined objects tending to occupy more eccentric orbits. For this high-order resonance, there are two different resonant modes. The principal one is the eccentricity-type mode, and we find that libration is permissible for orbits with $e\ge e_c^0$, where the critical eccentricity $e_c^0$ increases as a function of increasing inclination $i$. Correspondingly, we introduce a limiting curve $e_c^0(i)$, which puts constraints on the $(e, i)$ distribution of possible 4:7 resonators. We then perform numerical simulations on the sweep-up capture and long-term stability of the 4:7 MMR, and the results show that the simulated resonators are well-constrained by this theoretical limiting curve. The other 4:7 resonant mode is the mixed-$(e, i)$-type, and we show that stable resonators should exist at $i\gtrsim20^{\circ}$. We predict that the intrinsic number of these mixed-$(e, i)$-type resonators may provide a new clue into the Solar system's evolution, but, so far, only one real object has been observed resonating in this mode.

 \end{abstract}


\section{Introduction}

\begin{table}
\hspace{0cm}
\centering
\begin{minipage}{8.5cm}
\caption{There are 34 Kuiper belt objects that have been identified in the 4:7 MMR. These are three-opposition (or longer) observation arcs registered in the Minor Planet Center as of July 2018, with very small uncertainties in their measured semimajor axes ($\Delta a$/$a$) \citep{Chia2003, Glad2008}. The last column gives the orbital inclinations ($i$). The 4:7 RKBOs that were discussed by \citet{Lyka2007} are marked with an asterisk. 5 of 22 4:7 RKBOs in their list (1999 CD158, 1999 RH215, 2000 FX53, 2001 KP76, 2004 PW107) do not experience resonant angle libration for the full 10 Myr evolution, i.e., the longer observation arc available now compared to a decade ago shows that they are outside the 4:7 MMR. See the text for the definitions of the 0-mode and $-1$-mode.}      
\label{real4to7}
\begin{tabular}{l c c r}        

\bf{0-mode} \\   

\hline
\hline                 
~~~~~~~~~~~Object                    &      Oppositions         & $\Delta a$/$a$(\%)\footnote{The 1-$\sigma$ uncertainties $\Delta a$ in semimajor axis are taken from the ``Asteroids-Dynamic Site'' 
(https://newton.spacedys.com/astdys/)}    & $i$($^{\circ}$)             \\

\hline

~~~~~~~~~~~~~1999 HG12 (*)	           &	  6              &            0.16         &      1.7                                                \\

(129772) ~1999 HR11 (*)                &	  7              &            0.04         &      2.0                                                \\

(118378) ~1999 HT11  (*)                 &	  5              &            0.05        &      3.6                                                \\
                                  
~~~~~~~~~~~~~1999 KR18	 (*)                &	  3              &            0.09        &      2.2                                   \\    		

(60620) ~~2000 FD8	(*)                    &	   5              &           0.03        &     19.2                                                  \\     
                                       
~~~~~~~~~~~~~2000 OP67 (*)                &	   7              &           0.04      &      1.6                                                     \\                     

(118698) ~2000 OY51 (*)                 &	   5              &           0.05       &    12.8                                                    \\                
                   
(119066) ~2001 KJ76   (*)                   &	  4                &          0.05       &      6.1                                              \\            
                   
(160147) ~2001 KN76  (*)                 &	  8                &          0.03       &      4.0                                                  \\            
                   
(135024) ~2001 KO76 (*)                  &	   7               &          0.03        &     2.0                                                 \\            
                    
(119070) ~2001 KP77   (*)                 &	  4                &          0.04         &     3.6                                          \\            
                   
~~~~~~~~~~~~~2001 QE298 (*)                 &	      5            &          0.01       &      4.3                                     \\       
                   
(119956) ~2002 PA149  (*)                &	     7            &           0.03        &       2.4                                            \\   
                   
(135742) ~2002 PB171 (*)                 &	     4             &          0.07       &        6.5                                    \\   
                   
(385446) ~2003 QW111  (*)              &	    7              &          0.02       &       1.8                                    \\   
                   
~~~~~~~~~~~~~2003 QX91  (*)                &	     4             &          0.06        &        28.4                                       \\   
                   
(385527) ~2004 OK14  (*)               &	    4              &          0.03         &         2.5                                              \\   

(181871) ~1999 CO153	            &	  5              &            0.03         &      2.4                                     \\     
                   
(182222) ~2000 YU1                         &	   8               &           0.05       &          6.6                                      \\   
                   
~~~~~~~~~~~~~2004 SC60                     &	     5             &          0.02         &       2.1                                    \\          

~~~~~~~~~~~~~2004 VF131                      &	    6             &           0.04         &         2.2                          \\  
                  
(525816) ~2005 SF278                         &	    7             &           0.01          &          11.8                           \\           
   
~~~~~~~~~~~~~2012 YO9                           &	   4               &          0.08          &           13.9                           \\  
   
(531917) ~2013 BN82                        &	    6              &          0.03          &         6.7                           \\  
   
(532039) ~2013 FR28                         &	    4              &          0.02         &          2.8                             \\      
   
(500828) ~2013 GR136                      &	      4            &          0.01         &         2.3                       \\      
   
~~~~~~~~~~~~~2013 SB101                       &	      4            &          0.01         &         0.5                    \\      
   
~~~~~~~~~~~~~2013 UK17                      &	       3           &          0.02        &       25.0                                \\             
   
(533028) ~2014 AL55                       &	       5           &          0.01       &        2.7                                \\ 
   
~~~~~~~~~~~~~2014 UM229                  &	       4           &          0.01       &      5.5                         \\ 
   
~~~~~~~~~~~~~2015 BP518                   &	4                  &          0.01       &      8.6                                      \\    
   
~~~~~~~~~~~~~2015 BR518                   &	 3                 &          0.01        &     8.5                                    \\    
   
(536922) ~2015 FP345                      &	           3       &          0.02       &         8.6                                      \\    

\hline
\hline\\

\bf{$-1$-mode} \\

\hline
\hline                 
~~~~~~~~~~~Object                                 &      Oppositions         & $\Delta a$/$a$(\%) & $i$($^{\circ}$)  	 \\

\hline
  
(523742) ~2014	TZ85                        &          7                 & $<0.01$                &       16.0                                             \\

\hline
\hline
\end{tabular}
\end{minipage}
\end{table}

This paper extends our previous analysis of the high-inclination objects in the 2:3 (Plutinos) and 1:2 (Twotinos) mean motion resonances in the Kuiper belt \citep{Li2014a, Li2014b}. In the current work, we explore the 4:7 mean motion resonance (MMR) with Neptune, which is centered at a semimajor axis $a\approx43.8$ au. This high-order resonance is particularly interesting because it is near the dynamically cold ``kernel'' of the main classical Kuiper belt (MCKB). The MCKB extends between roughly 42 au $<a<$ 48 au  \citep{Lyka2005a, Glad2008, Peti2011}. Systematic studies of the 4:7 MMR were made a decade ago by \citet{Lyka2005b} and \citet{Lyka2007}, and they found that most stable orbits concentrate at low inclinations ($i<10^{\circ}$). The number of observed 4:7 resonant Kuiper belt objects (RKBOs) has doubled since then. As of July 2018, 34 samples have been identified as they show resonant angle libration\footnote{i.e., the resonant angle oscillates around the general libration center \citep{Li2014a, Li2014b} with an amplitude below $180^{\circ}$, rather than circulating through all possible angles $0^{\circ}-360^{\circ}$} for an entire 10~Myr orbital integration; and more high-inclination members with $i>10^{\circ}$ have been discovered (Table~\ref{real4to7}). Based on theoretical studies, the total number of 4:7 RKBOs with absolute magnitudes $H\le7$ is about $69\pm9$ \citep{adam2014}, while more recent observationally based estimates find that the $H_r\lesssim8.66$ population could be as large as 1000 \citep{volk2016}. Furthermore, \citet{Morb2014} showed that outward sweeping of the 4:7 MMR could reproduce the peculiar eccentricity distribution of the cold population in the MCKB, and the (high) inclinations of primordial planetesimals may play an important role. All these aspects motivate us to revisit the 4:7 MMR, with a particular focus on the high-inclination orbits with $i>10^{\circ}$.

\begin{figure}
 \hspace{0cm}
  \centering
  \includegraphics[width=9cm]{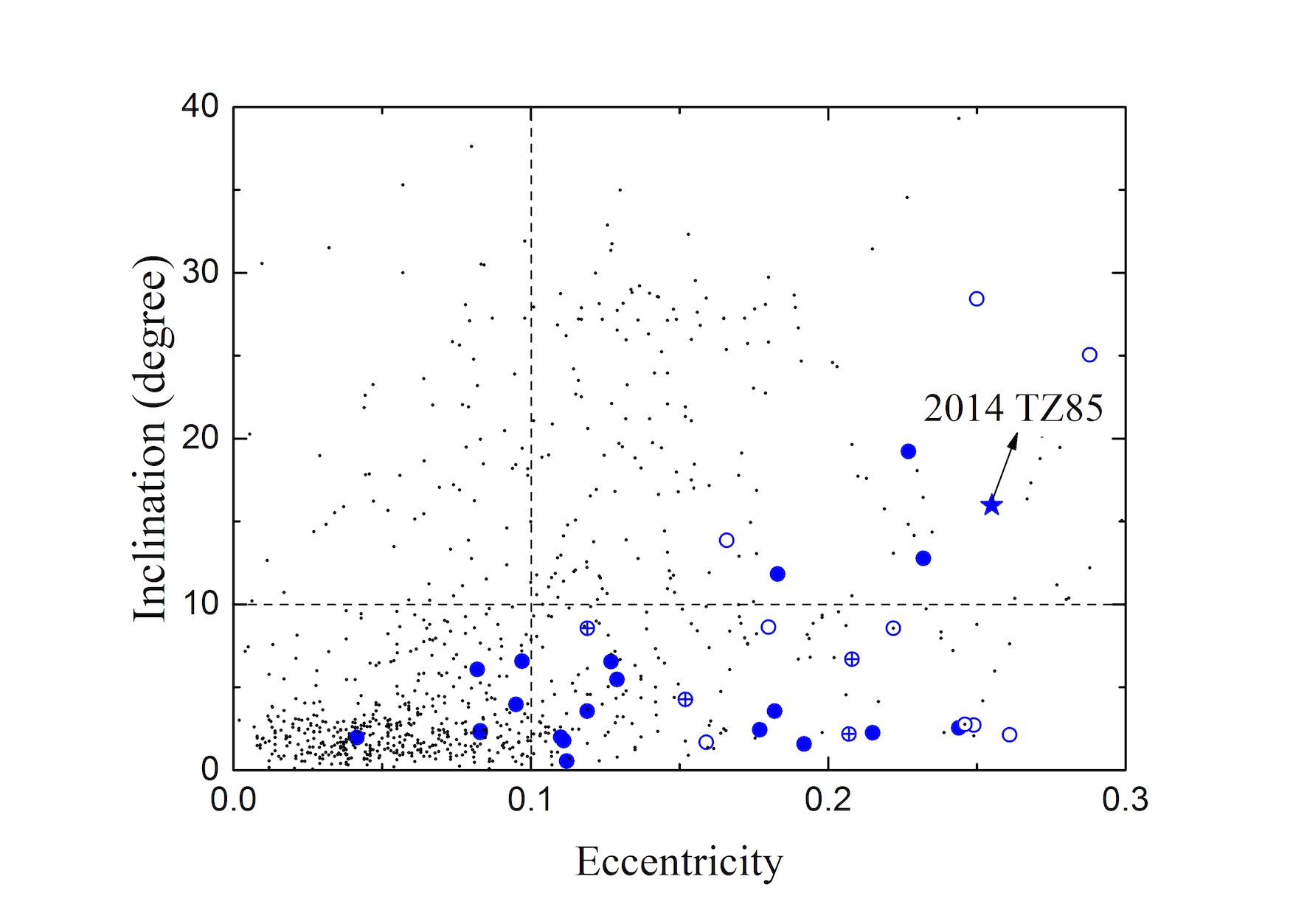}
  \caption{Distribution of osculating eccentricities and inclinations for the observed KBOs within the region of the MCKB. The blue symbols are the 4:7 RKBOs: the circles refer to the 0-mode resonators. Filled circles are only associated with the 4:7 MMR.  Open circles indicate those 4:7 resonators also experiencing the Kozai mechanism with the argument of perihelion $\omega$ centered at $0^{\circ}$ or 180$^{\circ}$, and circles with plus symbols inside have $\omega$ centered at 90$^{\circ}$ or 270$^{\circ}$; and the star refers to the single $-1$-mode resonator. The small black dots depict the other KBOs in the MCKB region but outside the 4:7 MMR. For reference, the vertical and horizontal dashed lines are plotted at $e=0.1$ and $i=10^{\circ}$, respectively.}
  \label{observed}
\end{figure}

Fig.~\ref{observed} shows the distribution of osculating eccentricities ($e$) and inclinations ($i$) for the Kuiper belt objects (KBOs) within the region of the MCKB, taken from the Minor Planet Center (MPC)\footnote{http://www.minorplanetcenter.net/iau/lists/TNOs.html}. Here the 4:7 RKBOs (as listed in Table~\ref{real4to7}) are highlighted by blue symbols. It can be seen that, for this resonant population, there is an apparent lack of high-inclination objects with $i>10^{\circ}$ in the low eccentricity range $e\lesssim0.1$; while objects can possess $i$ as high as $25^{\circ}$--$30^{\circ}$ on very eccentric orbits ($e\gtrsim0.25$). The known 4:7 RKBOs seem to have a tendency to occupy more inclined orbits at higher eccentricities. An important caveat to keep in mind is that the current census of the 4:7 RKBOs is strongly biased due to the observational incompleteness \citep[e.g. see discussion in][]{volk2016}. So, in order to know the true $(e, i)$ distribution, it is imperative to create a theoretical picture of the 4:7 MMR dynamics.

In previous works we studied the dynamics of high-inclination objects in the 2:3 and 1:2 MMRs \citep{Li2014a, Li2014b}, and we highlighted the distinctive resonant features for inclined orbits by defining the special libration center (SLC) of the resonant angle, which corresponds to the time variation of the semimajor axis $da/dt=0$. The mean value of the SLC is defined as the general libration center (GLC), which is usually called the ``resonant center''; the maximum excursion of the SLC from the GLC measures the lower limit of the resonant amplitude. For detailed descriptions and calculations of these parameters, the reader is referred to the two articles cited above.

The general resonant angle of an external $(p+q)$ : $p$ MMR is defined by
\begin{equation}
\sigma_j=p\lambda-(p+q)\lambda_N+(q-2j)\varpi+2j\Omega,
\label{angle}
\end{equation}
where $\lambda$, $\varpi$ and $\Omega$ are the mean longitude, the longitude of perihelion, and the longitude of ascending node, respectively; the subscript $N$ refers to Neptune and no subscript indicates the KBO; and $p(>0)$, $q(\le0)$, and $j$ are integers. On the right-hand side of Eq.~(\ref{angle}), the D'Alembert rules are obeyed as: (1) the sum of the coefficients of all the angles is equal to 0; (2) the sum of the coefficients of the longitudes of ascending nodes is an even number (i.e., the $2j$ coefficient of $\Omega$). The value of $|q|$ is referred to as the order of a resonance. The resonance's strength is proportional to $e^{|q-2j|}i^{|2j|}$, and it is straightforward to consider only the strongest case with the smallest $(|q-2j|+|2j|)$, corresponding to the lowest order of the resonant terms in the expansion of the disturbing function.

The first-order resonances (e.g., the 2:3 and 1:2 MMRs) only have a single resonant angle associated with the particle's eccentricity, while for the higher-order 4:7 MMR, there are two resonant angles to be taken into account:
\begin{equation}
\sigma_j=7\lambda-4\lambda_N-(3+2j)\varpi+2j\Omega,
\label{angle47}
\end{equation}
where $j=0$ and $-1$. Namely, one is the usual eccentricity-type resonance characterized by the angle
\begin{equation}
\sigma_0=7\lambda-4\lambda_N-3\varpi~~~~~~~~~~~~~(j=0),
\label{angle47e}
\end{equation}
and we refer to this as the ``0-mode'' in this paper. The other resonance is the mixed-$(e, i)$-type characterized by 
\begin{equation}
\sigma_{-1}=7\lambda-4\lambda_N-\varpi-2\Omega~~~~(j=-1),
\label{angle47i}
\end{equation}
correspondingly it will be referred as the ``$-1$-mode''. For the low-inclination population, the 0-mode dominates the 4:7 MMR and $\sigma_0$ librates; but for the high-inclination population, the independent $-1$-mode resonance may arise, i.e., $\sigma_{-1}$ librates while  $\sigma_0$ simultaneously circulates. Details of the individual modes will be addressed in the next section.

For the observed 4:7 RKBOs as shown in Table~\ref{real4to7}, the overwhelming majority are associated with the 0-mode (33 out of 34 in total). There is only a single object, 2014 TZ85, found to have $\sigma_{-1}$ librating independently. As we argued above, 2014 TZ85 indeed has a high inclination of $i\sim16^{\circ}$, and thus the relevant resonant strength ($\propto ei^2$) could be strong enough to maintain the $-1$-mode 4:7 resonance. 
We confirmed that it can maintain libration in the $-1$-mode for up to 1~Gyr. 

Given the long-term stability of 2014 TZ85, why is the $-1$-mode 4:7 resonance barely seen within the present-day MCKB? The answer to this question may depend on the resonance capture probability in individual modes during the outward migration of Neptune. The pronounced number difference between the 4:7 RKBOs currently librating in the 0-mode and $-1$-mode may place some constraints on the early dynamical evolution of the MCKB.

The outline of this paper is as follows: In Section 2 we reveal the role of $i$ on the 4:7 resonant feature using the semi-analytical method developed in \citet{Li2014a, Li2014b}. Consequently, we introduce a limiting curve in $(e, i)$ space which determines the permissible region of libration. In Section 3, focusing on the principal 0-mode of the 4:7 MMR, we numerically explore the long-term stability of high-inclination resonators, and verify the applicability of our theoretical limiting curve to predict the distribution of potential 4:7 RKBOs. In Section 4, we probe the effect of sweeping 4:7 resonance capture on test particles with different initial inclinations, and present the differences between the resulting 0-mode and $-1$-mode populations. Finally, the conclusions and discussion, including predictions for future observations, are given in Section 5.


\section[]{Resonant features and the permissible region}

\subsection{The 0-mode resonance}

\begin{figure}
 \centering
  \begin{minipage}[c]{0.5\textwidth}
  \hspace{-1 cm}
  \centering
  \includegraphics[width=8.6cm]{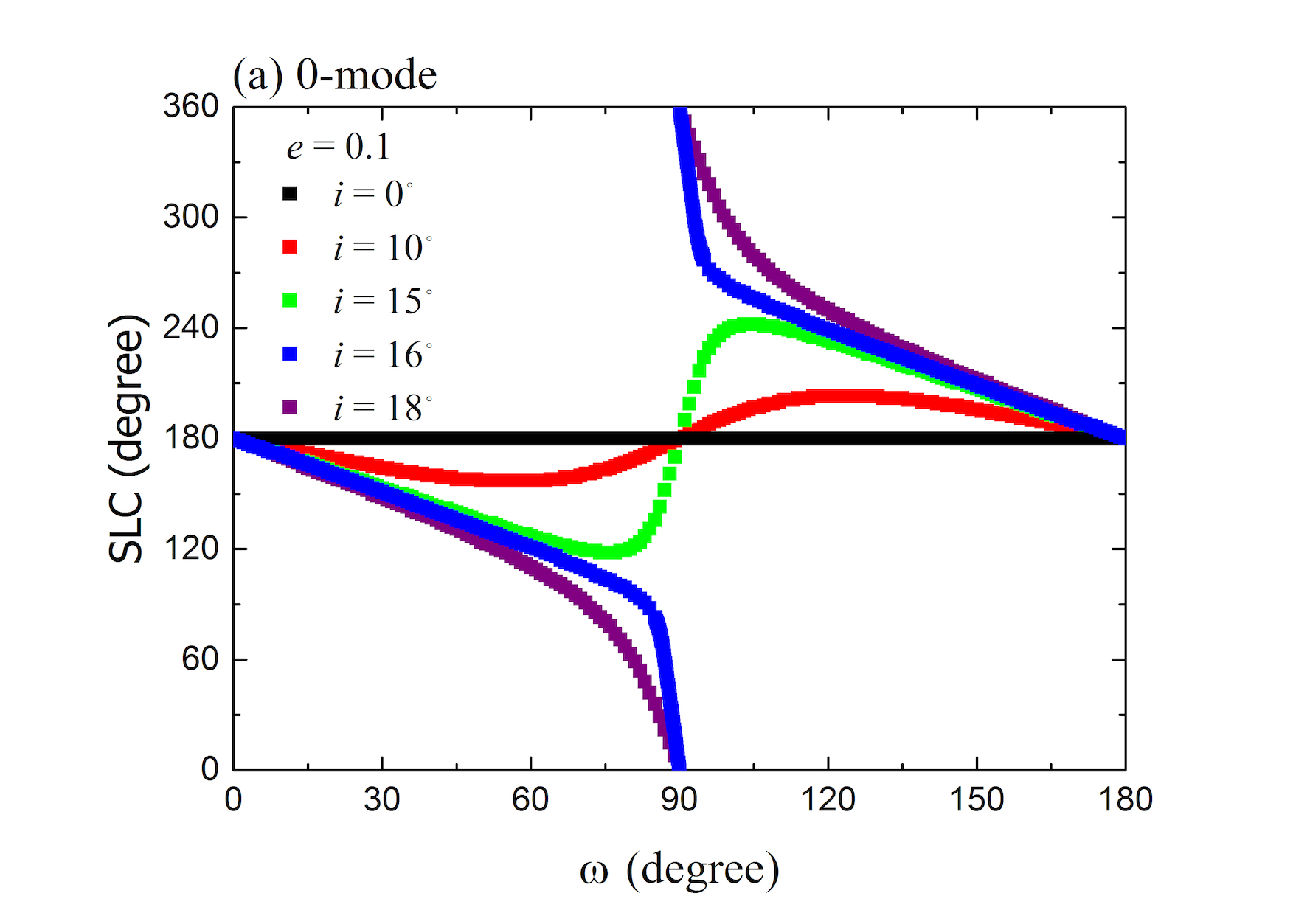}
  \end{minipage}
  \begin{minipage}[c]{0.5\textwidth}
  \hspace{-1 cm}
  \centering
  \includegraphics[width=8.6cm]{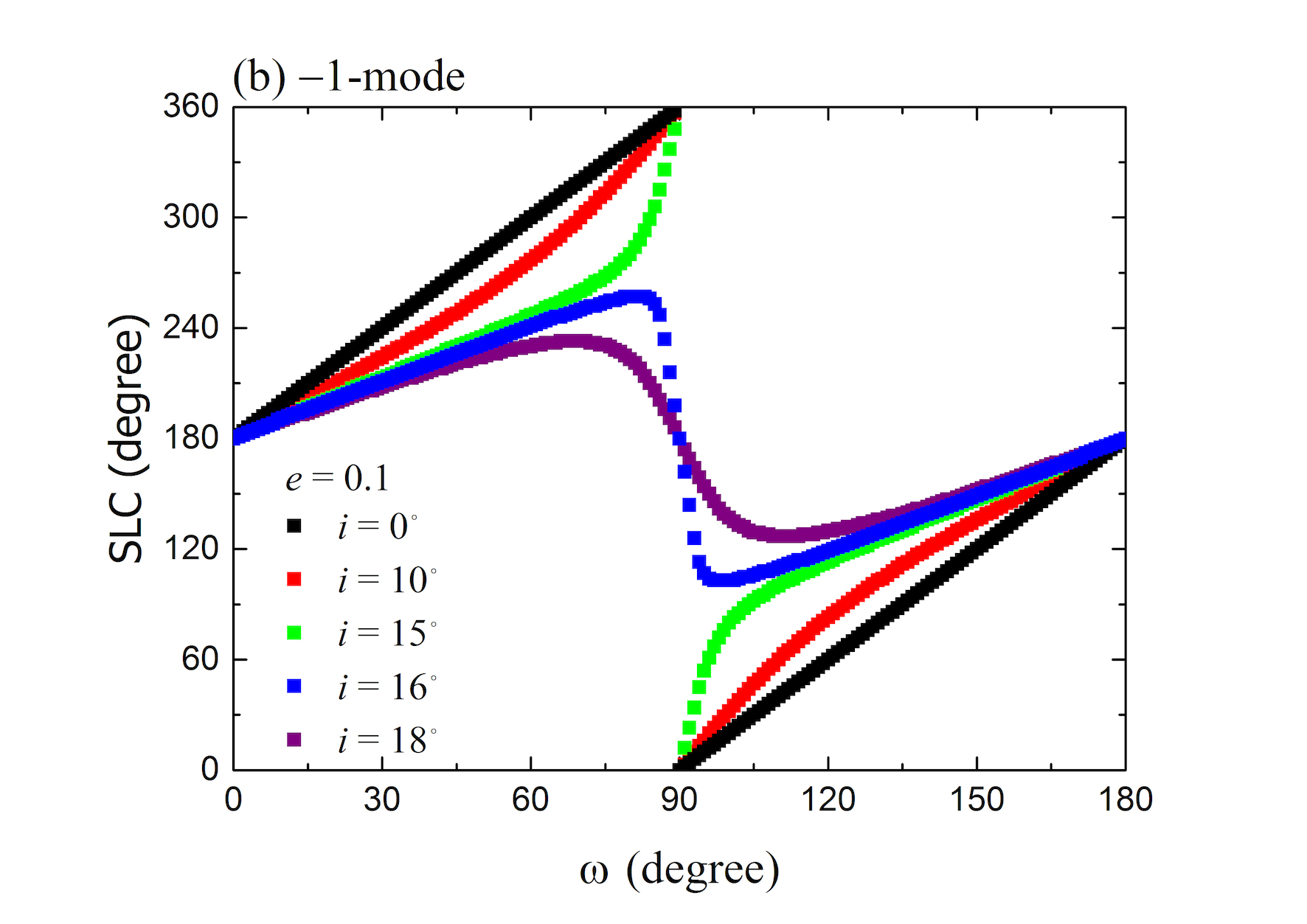}
  \end{minipage}
  \caption{The behavior of the SLC as the argument of perihelion $\omega$ varies between $0^{\circ}$ and $180^{\circ}$, with eccentricity $e$ fixed to be 0.1. (Panel a) For the 0-mode 4:7 MMR, at higher inclinations $i\ge16^{\circ}$ (blue and purple curves), the amplitude of the SLC can reach $180^{\circ}$, leading to circulation of the resonant angle $\sigma_0$. (Panel b) For the $-1$-mode 4:7 MMR, it is the lower inclinations $i\le15^{\circ}$ (black, red and green curves) where the SLC circulates, leading to circulation in the resonant angle $\sigma_{-1}$.}
  \label{slc}
\end{figure}

We first consider the 0-mode of the 4:7 MMR, which is associated with the resonant angle $\sigma_0$. The general behavior of this eccentricity-type resonance is similar to that of the 2:3 MMR \citep{Gall2006, Li2014a}: for high-inclination orbits, the SLC depends not only on the $(e, i)$ pair, but also strongly on the argument of perihelion $\omega$; and the average SLC corresponds to the libration center of $\sigma_0$ (i.e., GLC), which is fixed at $180^{\circ}$. With respect to the 2:3 MMR, we further pointed out that the amplitude of the SLC increases with larger $i$ ($\le90^{\circ}$) and smaller $e$ ($\le0.3$), reaching a maximum value of $\sim75^{\circ}$. Since the amplitude of the SLC defines the lower limit of the resonant amplitude $A_{\sigma}$, libration (i.e., $A_{\sigma}<180^{\circ}$) is always possible.

However, the specific features of the 4:7 MMR can be qualitatively different from the 2:3 MMR. We find that, for this high-order resonance, the SLC may eventually reach a transition from libration to circulation when $i$ is large enough. Fig.~\ref{slc}(a) shows that, for fixed $e=0.1$, the motion of the SLC is bounded at $i\le15^{\circ}$ and constrained to an amplitude below $\sim60^{\circ}$; but for  $i\ge16^{\circ}$, the motion of the SLC becomes unbounded and can circulate through the entire range of $0^{\circ}-360^{\circ}$. Because the amplitude of the SLC is the lower limit of the resonant amplitude $A_{\sigma_0}$, it implies that the particles with $i\ge16^{\circ}$ must have $A_{\sigma_0}=180^{\circ}$, and so would be prohibited from libration in the 4:7 MMR. Therefore, given $e=0.1$, the highest $i$ of the 0-mode 4:7 resonators may be around $15^{\circ}$. This mechanism could result in the lack of high-inclination 4:7 RKBOs in the region of $e\lesssim0.1$, as shown in Fig.~\ref{observed}. This effect is more extreme for lower values of $e$, for instance, if $e=0.05$, then the inclination of a librating 0-mode 4:7 RKBO should be no higher than $\sim7^{\circ}$.

\begin{figure}
 \hspace{-0.7cm}
  \centering
  \includegraphics[width=9cm]{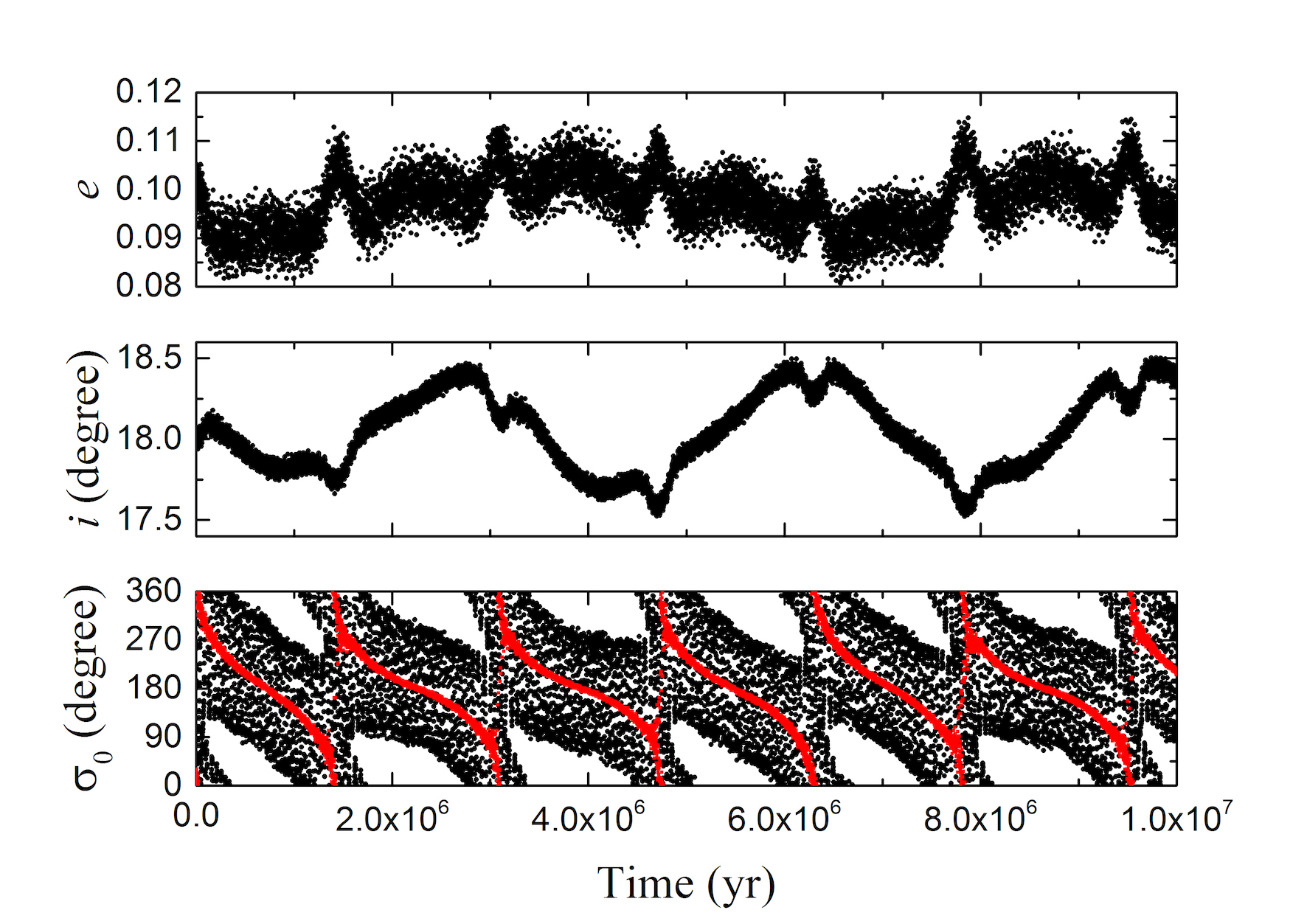}
  \caption{Time evolution of the eccentricity $e$, the inclination $i$ and the 0-mode resonant angle $\sigma_0=7\lambda-4\lambda_N-3\varpi$ for a particle with initial $e=0.1$ and $i=18^{\circ}$, starting at $\sigma_0=180^{\circ}$. The red curve in the bottom panel indicates the temporal variation of the SLC, circulating between $0^{\circ}$ and $360^{\circ}$, causing circulation of $\sigma_0$, thus making this particle non-resonant.}
  \label{circulation}
\end{figure}

Using a direct numerical simulation, in Fig.~\ref{circulation} we show the evolution of an inclined particle started at the 4:7 GLC: $\sigma_0=180^{\circ}$ with an initial $e=0.1$ and $i=18^{\circ}$, under the gravitational effects of the four Jovian planets\footnote{In this paper, to perform all numerical calculations, we use the SWIFT\_RMVS3 symplectic integrator developed by \citet{Levi1994}. We adopt a time-step of 0.5 yr, which is about 1/12 of the shortest orbital period (Jupiter's) in our simulations.}. From the bottom panel, one can easily see that the secular behavior of $\sigma_0$ indeed matches the temporal variation of the SLC (red curve). Since the SLC circulates, the resonant amplitude $A_{\sigma_0}$ is forced to reach $180^{\circ}$, thus making the particle non-resonant. A detailed analysis was made in \citet{Li2014a}.
Here, both $e$ and $i$ of the particle exhibit only very small oscillations during the entire integration. This is crucial for determining the SLC as we will demonstrate next.

\begin{figure}
 \centering
  \begin{minipage}[c]{0.5\textwidth}
  \hspace{-1 cm}
  \centering
  \includegraphics[width=9cm]{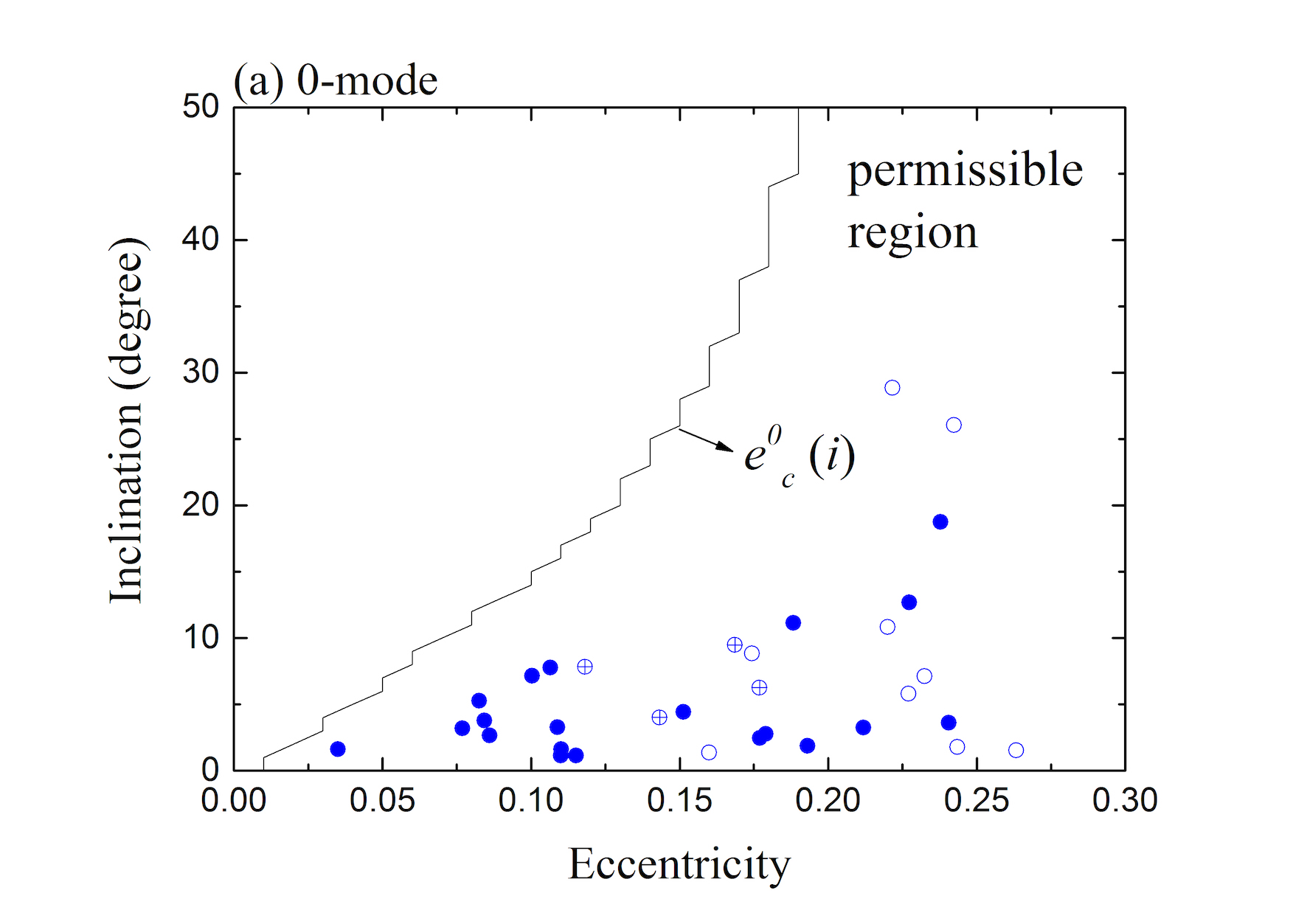}
  \end{minipage}
  \begin{minipage}[c]{0.5\textwidth}
  \hspace{-1 cm}
  \centering
  \includegraphics[width=9cm]{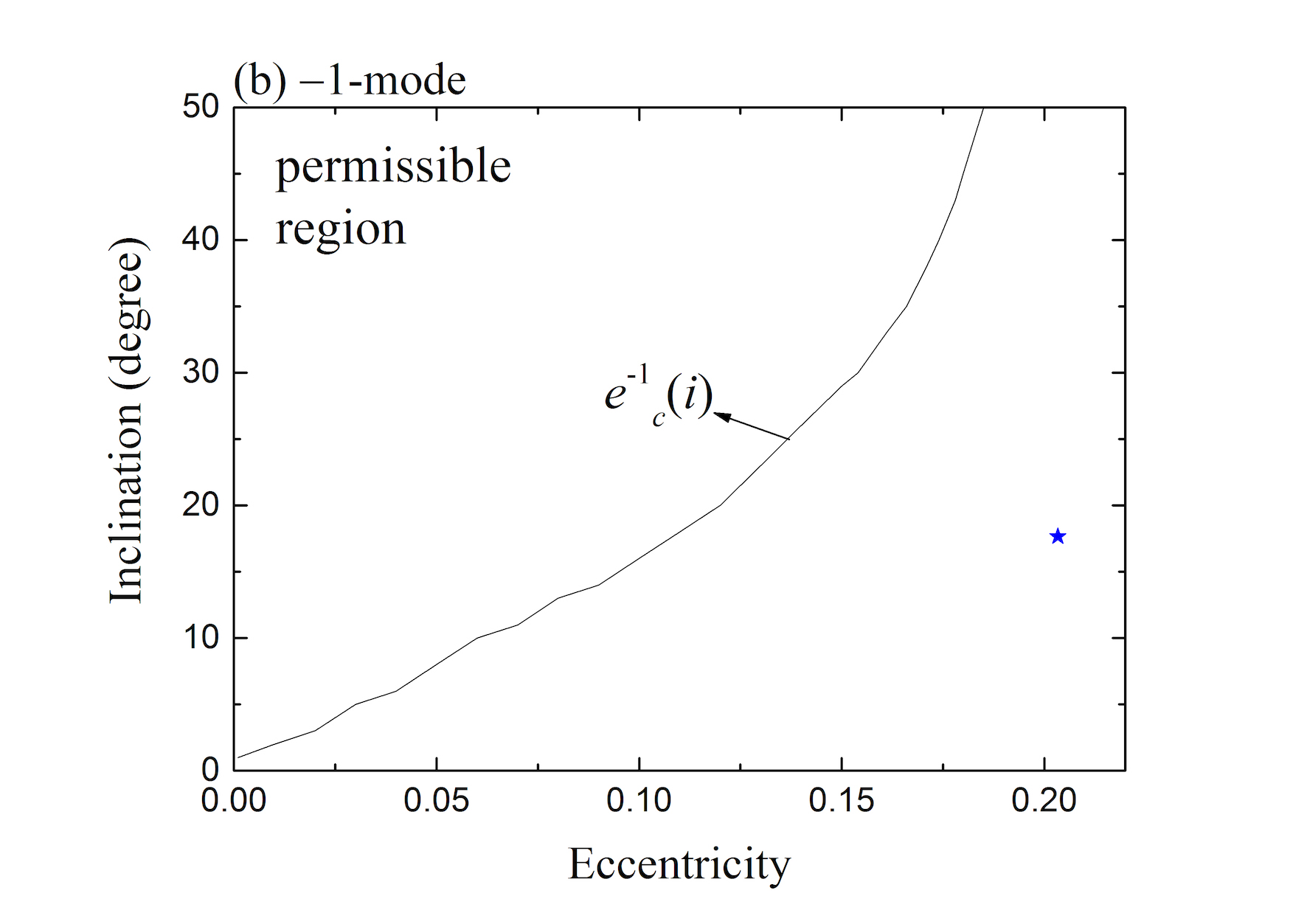}
  \end{minipage}
  \caption{(Panel a) For the 0-mode 4:7 MMR, the black curve indicates the critical eccentricity $e_c^0$ against the inclination $i$, i.e., the limiting curve $e_c^0(i)$. The permissible region of this resonance is on the right side, where all the known 4:7 RKBOs reside. The meanings of blue symbols are the same as in Fig.~\ref{observed}. For consistency with the calculation of $e_c^0$, the eccentricities and inclinations of the 4:7 RKBOs are taken to be the average values over the 10 Myr integration. (Panel b) For the $-1$-mode 4:7 MMR, the permissible region lies to the left of the corresponding limiting curve $e_c^{-1}(i)$. The star indicates the single observed sample 2014 TZ85, which is outside the permissible region due to the Kozai mechanism (see Section~2.2).}
  \label{PR}
\end{figure}

Analogous to the 2:3 MMR, for a given high value of $i$, the amplitude of the 4:7 SLC decreases with increasing $e$. We find that there is a critical eccentricity $e_c^0$: when $e\ge e_c^0$, the SLC will switch from circulation to libration. In other words, only if this $e$-condition is fulfilled, the resonant angle $\sigma_0$ is allowed to librate. We also find that, for the 0-mode 4:7 MMR, this critical eccentricity $e_c^0$ increases as a function of $i$. Fig.~\ref{PR}(a) gives the values of $e_c^0(i)$ derived from the libration/circulation behavior of the SLC for orbits with $i$ up to $50^{\circ}$ (black curve). This curve of the critical eccentricity as a function of inclination, $e_c^0(i)$, will be referred to as the ``limiting curve'' hereafter. The permissible region, where the 0-mode 4:7 resonators could librate, lies to the right of the limiting curve. 

Next, we would like to explore the $(e, i)$ distribution of the observed 4:7 RKBOs by comparing with the limiting curve. For the above semi-analytical method to obtain the limiting curve $e_c^0(i)$, we use the semimajor axis $a$ corresponding to the nominal position of the 4:7 MMR ($\sim43.8$ au), and for a specific pair of ($e, i$). But the numerically identified 4:7 RKBOs actually have orbital elements $(a, e, i)$ that evolve over time. In order to place an object approximately at the nominal resonance, as \citet{Lyka2005b} did, we compute its proper orbital elements by averaging the 2000 yr timestep output from the entire 10 Myr integration. These proper elements for the 4:7 RKBOs in the 0-mode are added to Fig.~\ref{PR}(a) in order to be consistent with the limiting curve calculation. The most important result of this paper is shown in Fig.~\ref{PR}(a): all of the real 0-mode 4:7 RKBOs are distributed in the permissible $(e, i)$ region to the right of the limiting curve, as predicted by our theory, with higher inclinations allowed at higher eccentricities.

We need to stress that our semi-analytical approach only takes into account the MMR, inside which the argument of perihelion $\omega$ is assumed to circulate between $0^{\circ}$ and $360^{\circ}$. The Kozai mechanism \citep{kozai62} causes $\omega$ to librate around $0^{\circ}$, $90^{\circ}$, $180^{\circ}$, or $270^{\circ}$, instead of circulating, and can occur at moderate inclinations inside MMRs \citep{kozai85}. We will refer to RKBOs with different Kozai centers as $K_X$-type, where $X$ is one of the four possible libration centers of $\omega$. If the Kozai mechanism is dominant, the libration of $\omega$ may narrow the variation of the SLC, and consequently affect the validity of our limiting curve for constraining the $(e, i)$ space of the 4:7 resonators. Libration of the resonant angle $\sigma_0$ will be influenced differently depending on the Kozai center: \\
$~~~\bullet$ $K_{90}$-type and $K_{270}$-type 4:7 Kozai resonators: Referring back to Fig.~\ref{slc}(a), one can see that orbits with $e<e_c^0(i)$ (blue and purple curves) have the SLCs crossing $0^{\circ}$ (equivalent to $360^{\circ}$) at $\omega=90^{\circ}$; and have a similar behavior at $\omega=270^{\circ}$.
But because the libration center of the 4:7 resonance is located at $\sigma_0=180^{\circ}$, the SLC must also pass the value of $180^{\circ}$. These effects in combination mean that even though $\omega$ is confined by the Kozai mechanism, when $e<e_c^0(i)$, the SLC will still traverse from $0^{\circ}$, through $180^{\circ}$ and toward $360^{\circ}$, resulting in circulation of $\sigma_0$. This can be seen in Fig.~\ref{PR}(a): all of the 0-mode 4:7 RKBOs involved in the $K_{90}$- or $K_{270}$-type Kozai mechanism (blue circles with plus symbols inside), fall into the permissible region.\\
$~~~\bullet$ $K_{0}$-type and $K_{180}$-type 4:7 Kozai resonators: Again referring to Fig.~\ref{slc}(a), it shows that when $\omega$ is librating around $0^{\circ}$ or $180^{\circ}$, the SLC will be confined to angles near $180^{\circ}$, regardless of whether or not the eccentricity is below the critical value $e_c^0$. As indicated by the blue open circles in Fig.~\ref{PR}(a), as long as the eccentricities are sufficiently large ($e>0.15$), the 4:7 RKBOs can possibly experience the $K_{0}$- or $K_{180}$-type Kozai mechanism not only at high inclinations but also at very small inclinations ($i<2^{\circ}$). Although none of the observed 0-mode 4:7 RKBOs violate the constraint of the limiting curve, possible exceptions with $e<e_c^0(i)$ may arise from a later stability study of the 4:7 MMR, but they are apparently rare.  

We note that with the Kozai mechanism in effect, there could be correlation between the two resonant angles:
\begin{equation}
\sigma_{-1}=\sigma_0+2\omega,
\label{kozai}
\end{equation}
here the angle $\sigma_{-1}$ could also be librating about the resonant center at $180^{\circ}$. We classify this type of coupled resonance into the 0-mode, which is supposed to dominate the 4:7 MMR behavior.

\subsection{The $-1$-mode resonance}

\begin{figure}
 \hspace{-0.7 cm}
  \centering
  \includegraphics[width=9cm]{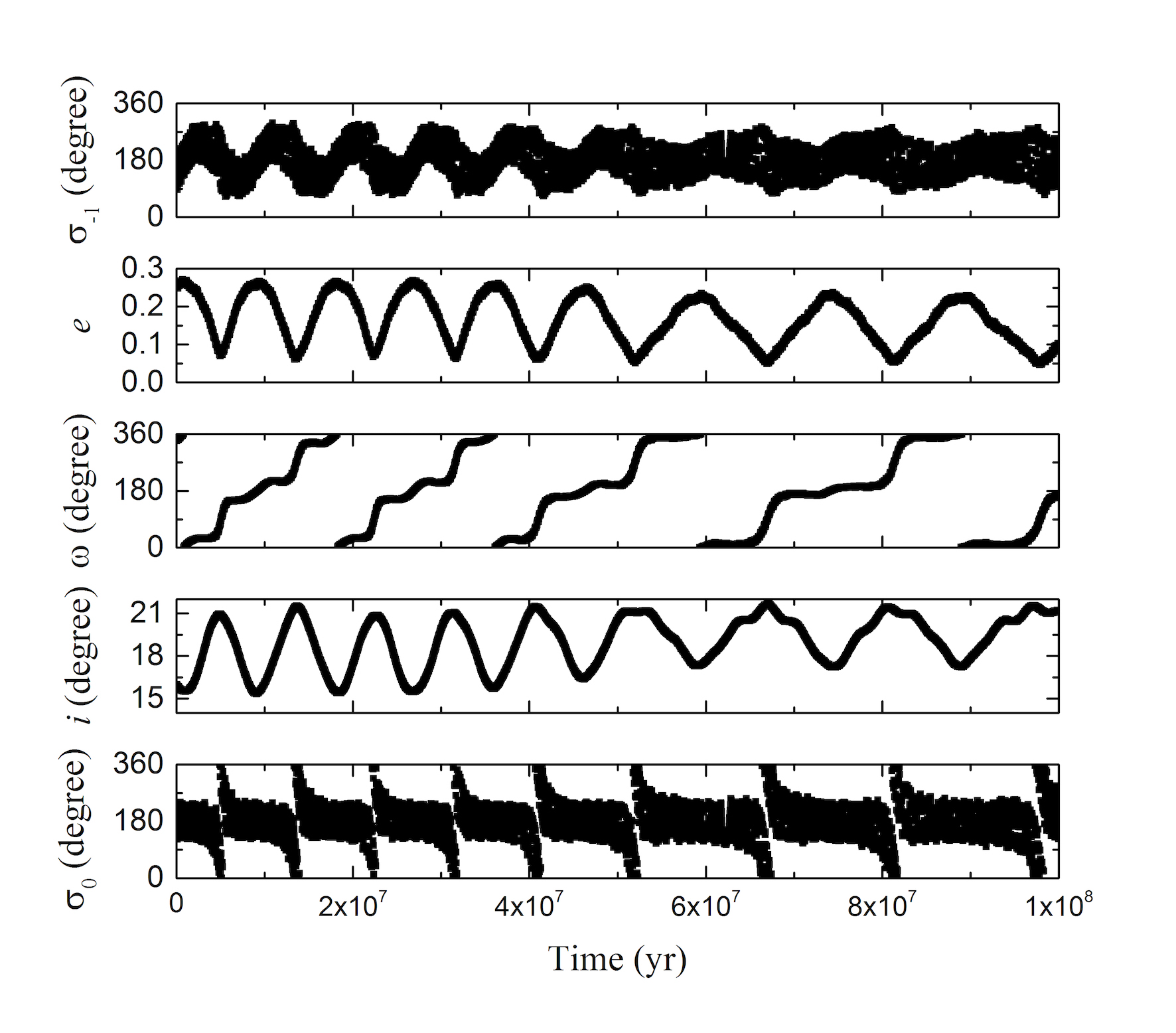}
  \caption{For the observed KBO 2014 TZ85 trapped in the independent $-1$-mode of the 4:7 MMR, the time evolution of the $-1$-mode resonant angle $\sigma_{-1}=7\lambda-4\lambda_N-\varpi-2\Omega$, the eccentricity $e$, the argument of perihelion $\omega$, the inclination $i$, and the 0-mode resonant angle $\sigma_0$ (top to bottom).}
  \label{2014TZ85}
\end{figure}

When we refer to the $-1$-mode of the 4:7 MMR, we mean the independent state where $\sigma_{-1}$ is librating while $\sigma_0$ is circulating. Fig.~\ref{slc}(b) shows the behavior of the $-1$-mode SLC for orbits with $e=0.1$. The unbounded motion takes place at $i$ smaller than $15^{\circ}$, resulting in the circulation of $\sigma_{-1}$. Then similarly, in Fig.~\ref{PR}(b) we plot the limiting curve $e_c^{-1}(i)$ for the $-1$-mode resonance, and the corresponding permissible region is on the left side, i.e., requiring $e\le e_c^{-1}(i)$ for libration. Considering the resonance's strength is proportional to $ei^2$, it is likely that the 4:7 resonators could be in the $-1$-mode on either high-$i$ or small-$i$-and-large-$e$ orbits, but the profile of the limiting curve $e_c^{-1}(i)$ seems to favor the former. We further note that, given $e>0.19$, the SLC of the $-1$-mode resonance always varies from $0^{\circ}$ to $360^{\circ}$ for any $i\le50^{\circ}$, leading to the persistent circulation of $\sigma_{-1}$. This means that the 4:7 RKBOs with $e>0.19$ should not be resonating in the $-1$-mode.

Unexpectedly, the only $-1$-mode resonator (2014 TZ85) among the observed 4:7 RKBOs is outside the permissible region (see Fig.~\ref{PR}(b)). Examining its orbital behavior in Fig.~\ref{2014TZ85}, one can see the resonant angle $\sigma_{-1}$ is librating even when $e>e_c^{-1}$. This libration is because of the influence of the Kozai mechanism, as we discussed for the 0-mode resonance above: if a KBO has its $\omega$ librating around $0^{\circ}$ or $180^{\circ}$, the $-1$-mode SLC can librate even if the condition $e\le e_c^{-1}(i)$ is not fulfilled (see Fig.~\ref{slc}(b)). Consequently, during the high-$e$ stage, the libration of $\sigma_{-1}$ is possible. We note that $\omega$ does not strictly librate in the integration, but appears to librate about a circulating center. This minor $\omega$-libration seems to be enough to affect the behavior of $\sigma_{-1}$.
Here, the $-1$-mode resonance is the dominant type for 2014 TZ85, since the 0-mode resonant angle $\sigma_0$ is circulating, as shown in the bottom panel of Fig.~\ref{2014TZ85}.

Due to the extreme rarity of observed $-1$-mode 4:7 librators, we think it is not necessary to explore the long-term stability of the $-1$-mode resonance in a fictitious $(e, i)$ space. The fact that we have detected so few $-1$-mode resonators could be a result of the sweeping 4:7 MMR capture process, and may thereby provide some clues to the primordial distribution of planetesimals in the MCKB. This will be discussed further in the planet migration and resonance capture simulations in Section 4.


\section{Dynamical evolution} \label{sec:dyn}

Having found the peculiar features of the 4:7 MMR, we now present a detailed analysis of numerical simulations within this resonance. We note for the reader that there is a summary of the main points covered here in Section~\ref{sec:dynsum}.

In this section we will consider only the principal (and most common) 0-mode population in the 4:7 MMR. In doing so, we intend to explore how the potential 4:7 RKBOs are distributed in $(e, i)$ space and whether we should expect that they could fill the entire permissible region shown in Fig.~\ref{PR}(a). The numerical experiments conducted by \citet{Lyka2005b} suggest that resonant objects may suffer chaotic diffusion and have very different initial and final values of both $e$ and $i$ after Gyrs of dynamical evolution. Nevertheless, the analysis above implies that the constraint of our limiting curve should always stand.  

Similar to what we did in \citet{Li2014b}, in pre-runs we firstly derive the initial conditions for resonant particles in the framework of the present outer Solar system. We choose the initial orbital element space to be $(e=0.1, i=10^{\circ}-20^{\circ}, \Delta i=1^{\circ})$, $(e=0.15, 0.2, 0.25, 0.3,  i=10^{\circ}-30^{\circ}, \Delta i=5^{\circ})$. For each set of initial $(e, i)$, we start with 2001 massless particles uniformly placed in the $a$ space between 42.5 and 44.5 au, with a high resolution of $\Delta a=0.001$ au. Here and in the next subsection, the initial angles are chosen such that all particles begin with $\sigma_0=180^{\circ}$, $\omega=90^{\circ}$ and $\Omega=0^{\circ}$. We then numerically compute the resonant amplitudes $A_{\sigma_0}$ over the 10 Myr integration, which is the same as the timespan used for identifying the real 4:7 RKBOs.

\subsection{$i$-dependent boundaries}

\begin{figure}
 \centering
  \begin{minipage}[c]{0.5\textwidth}
  \hspace{-1 cm}
  \centering
  \includegraphics[width=9cm]{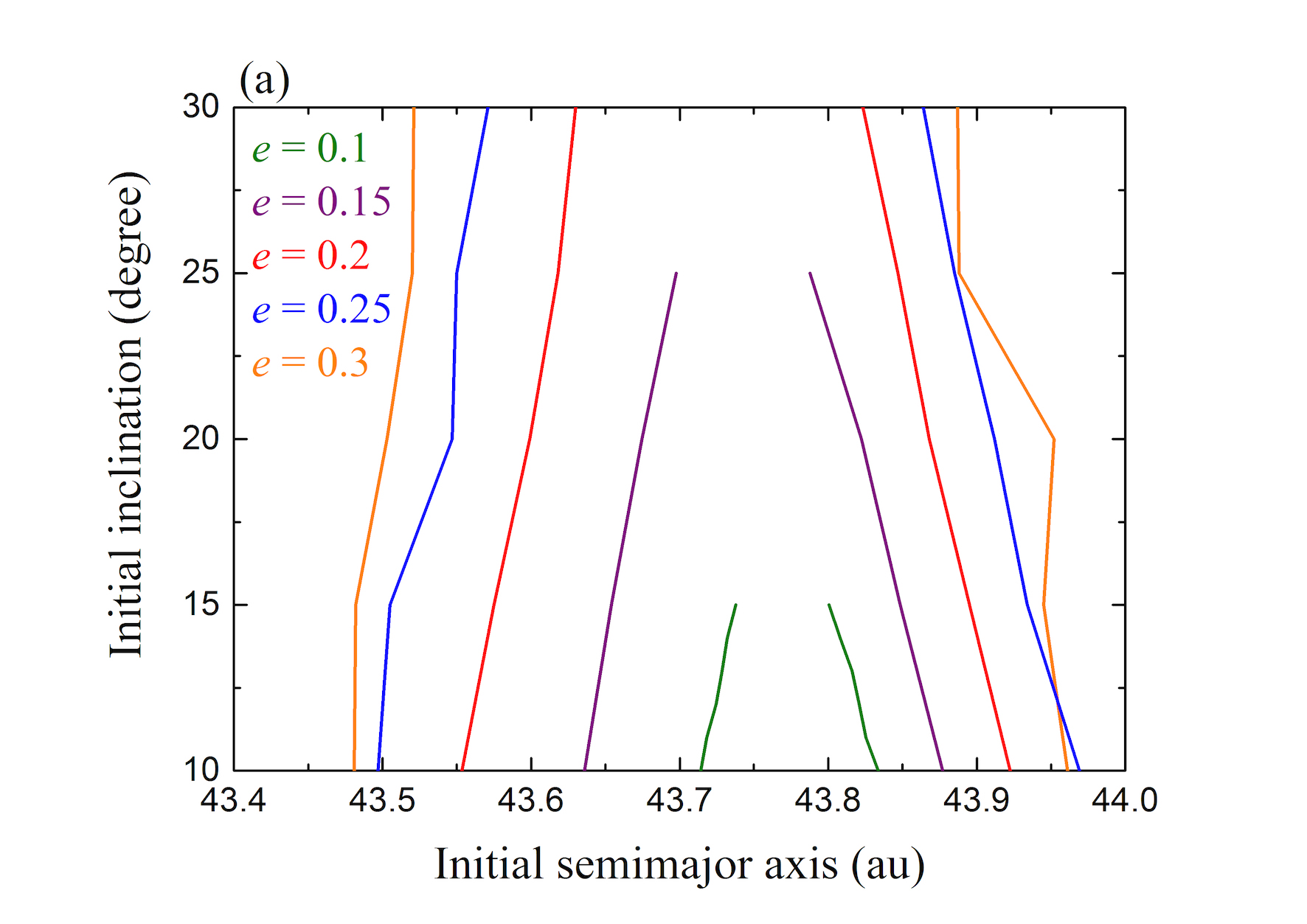}
  \end{minipage}
  \begin{minipage}[c]{0.5\textwidth}
  \hspace{-1 cm}
  \centering
  \includegraphics[width=9cm]{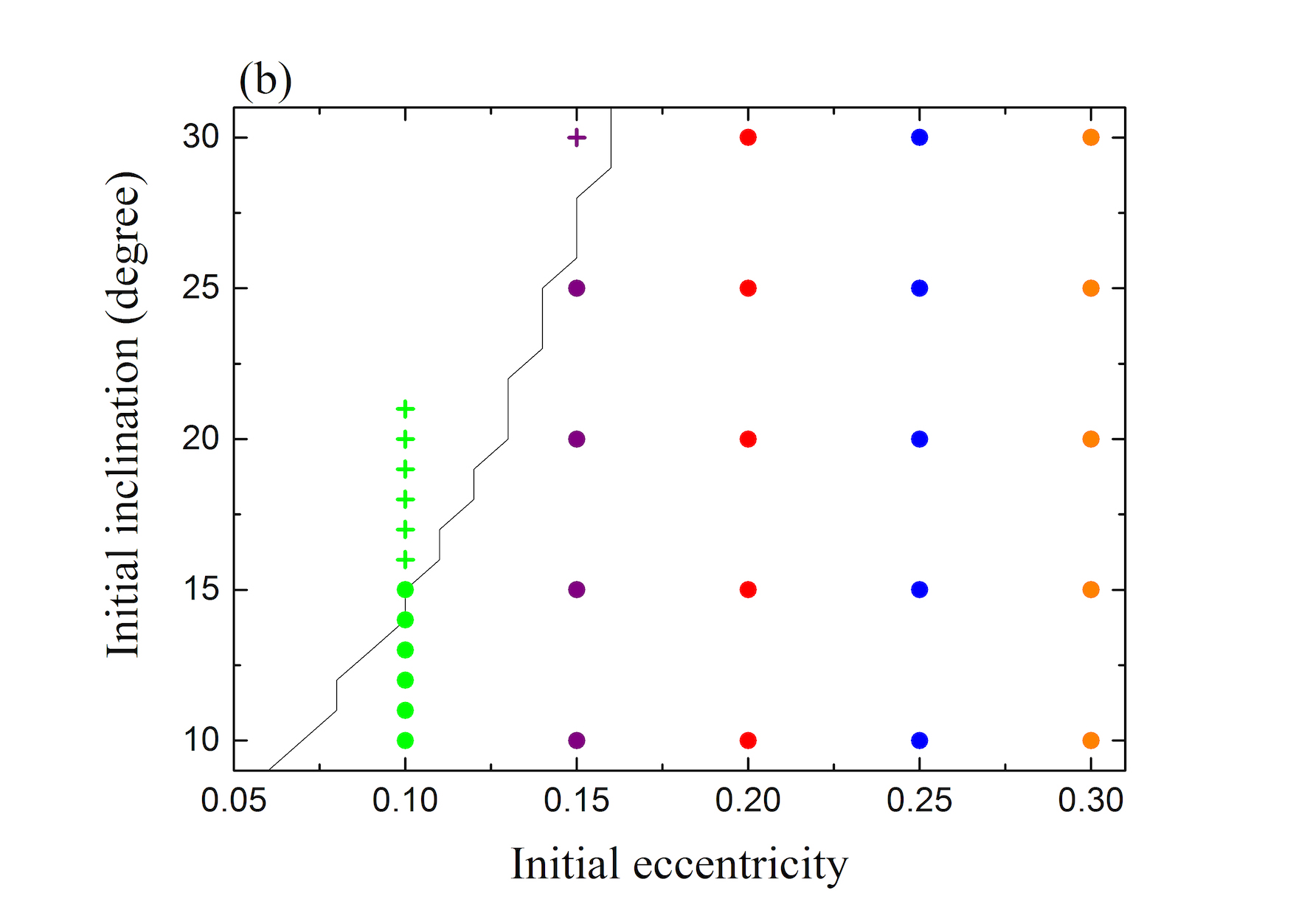}
  \end{minipage}
  \caption{(Panel a) The boundaries of the 0-mode 4:7 MMR on the initial $(a, i)$ plane for different initial $e$. Given a relatively small $e=0.1$ (0.15), the green (purple) curves are cut off at $i=15^{\circ}$ ($25^{\circ}$), above which the libration motion of $\sigma_0$ is not possible; (Panel b) Corresponding distribution of the librating (dots) and circulating (crosses) particles on the initial $(e, i)$ plane. The limiting curve is represented by the black line.}
  \label{boundary}
\end{figure}

\begin{figure*}
  \centering
  \begin{minipage}[c]{1\textwidth}
  \vspace{0 cm}
  \includegraphics[width=9cm]{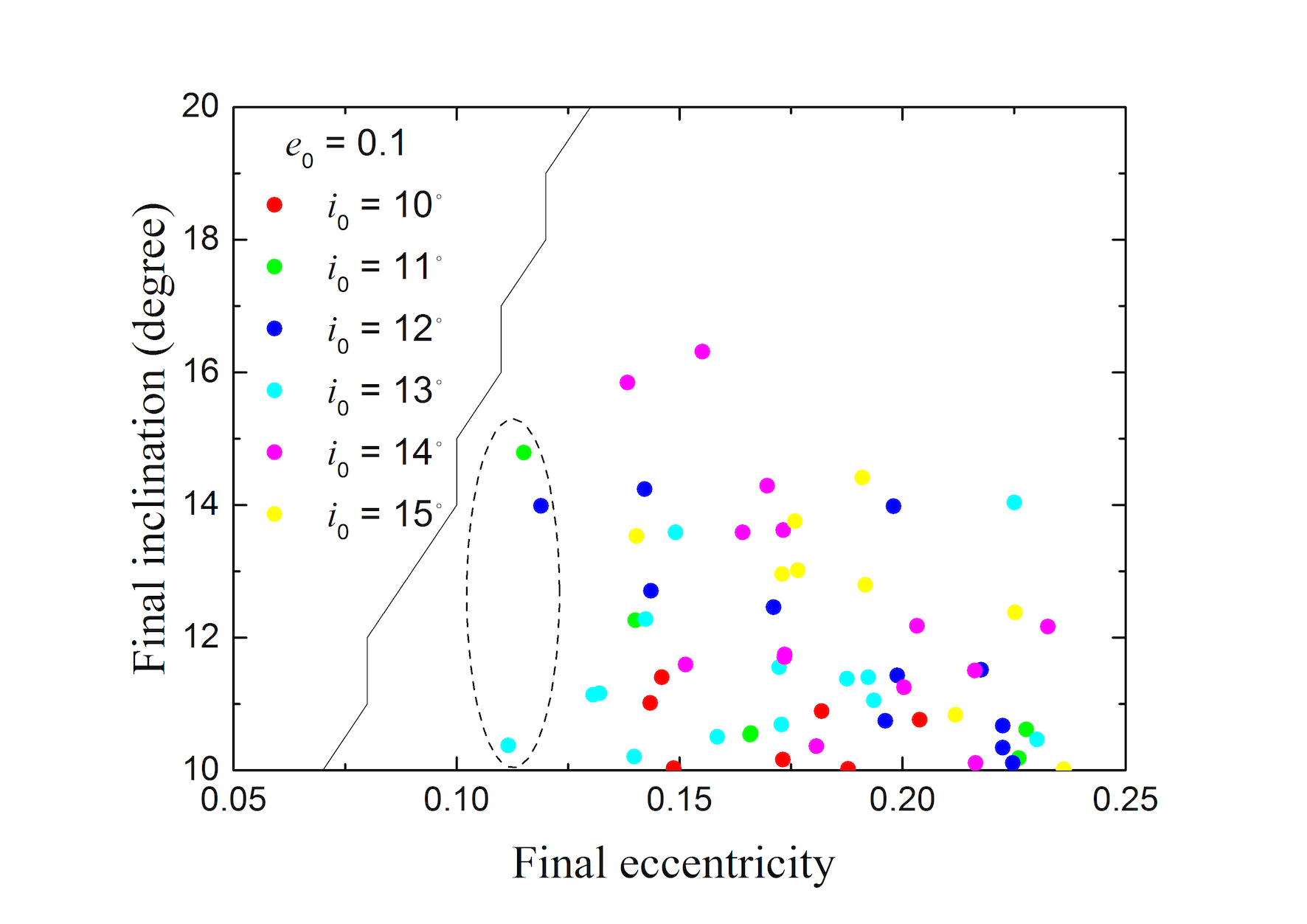}
  \includegraphics[width=9cm]{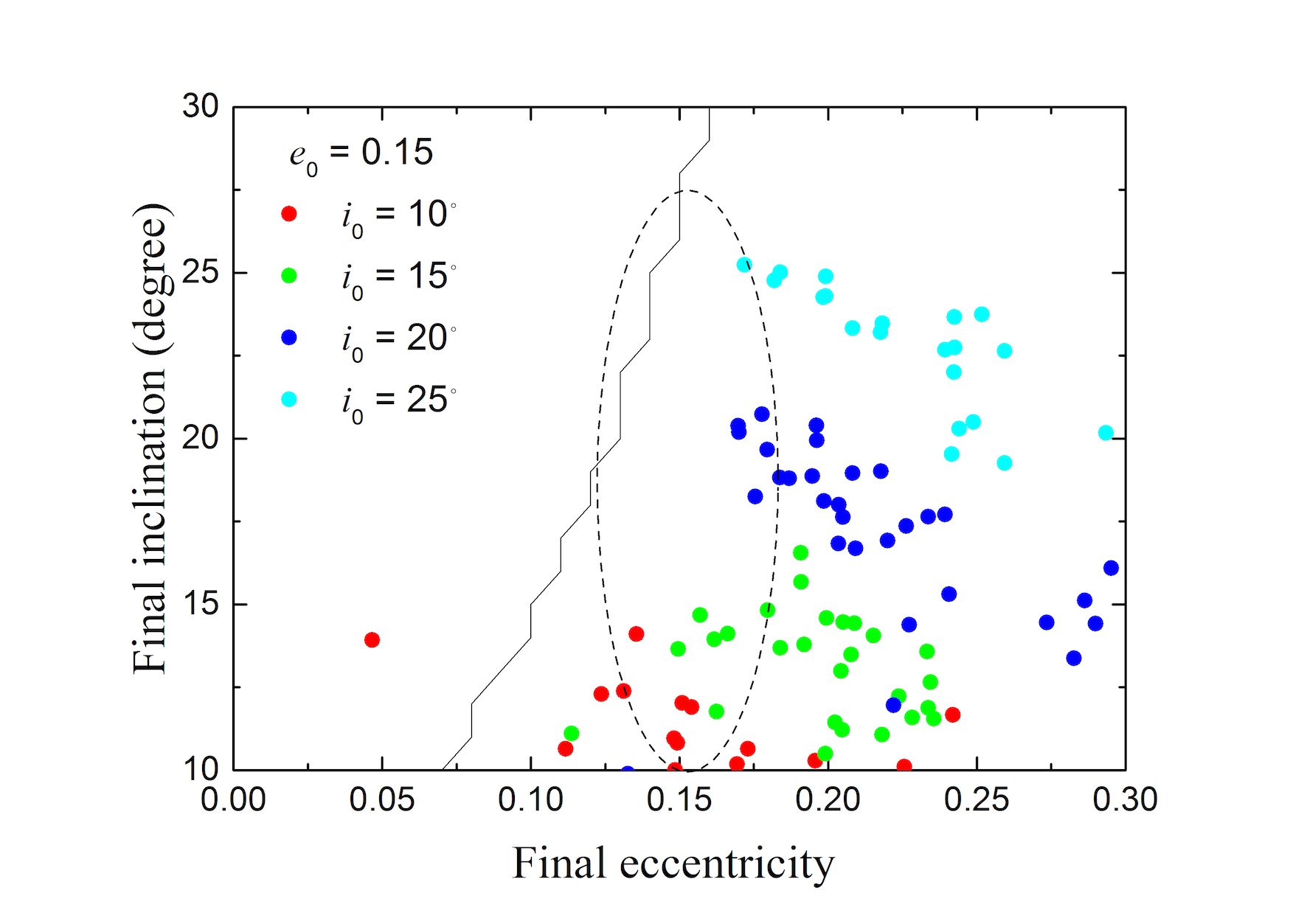}
  \end{minipage}
  \begin{minipage}[c]{1\textwidth}
  \vspace{0 cm}
  \includegraphics[width=9cm]{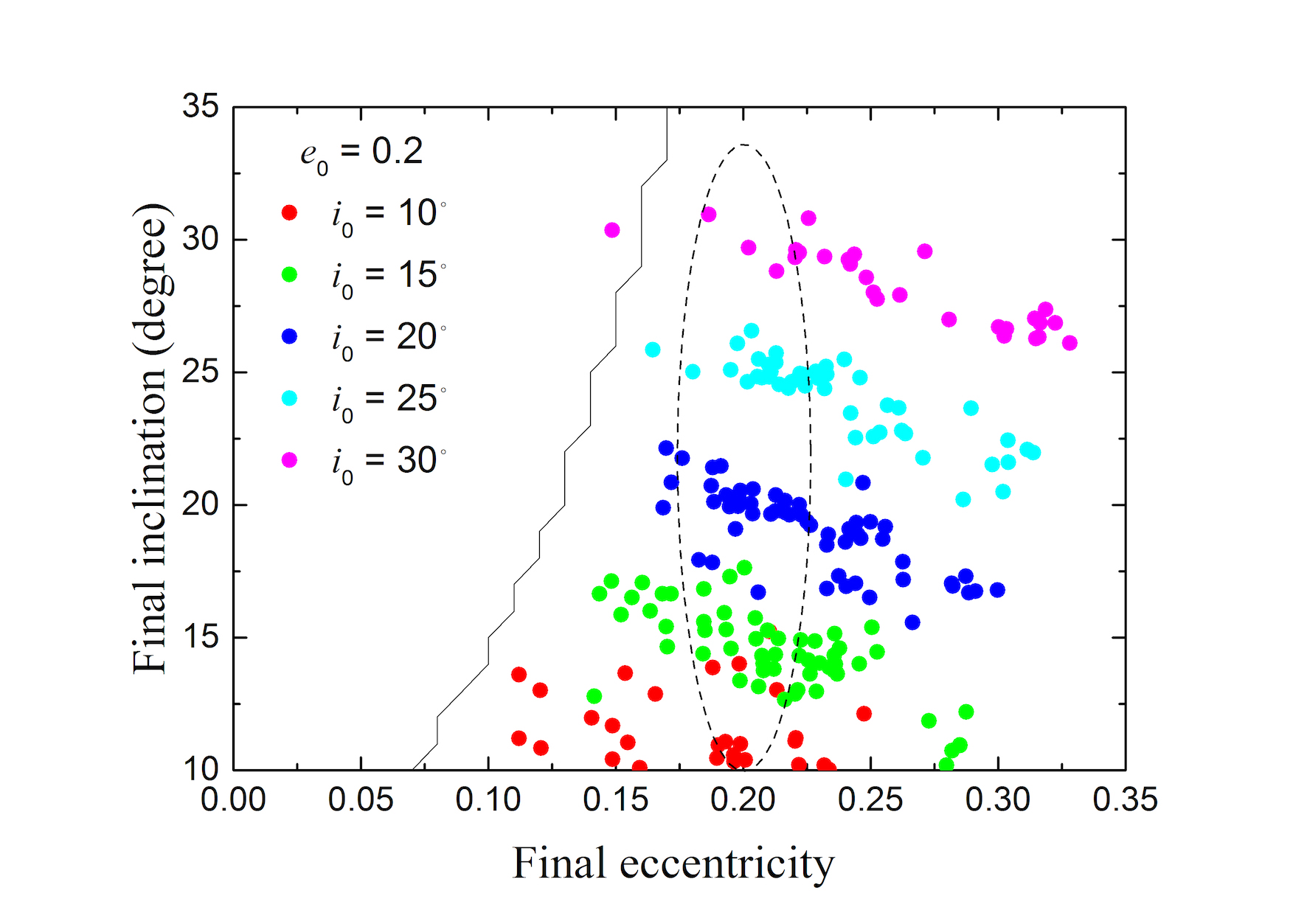}
  \includegraphics[width=9cm]{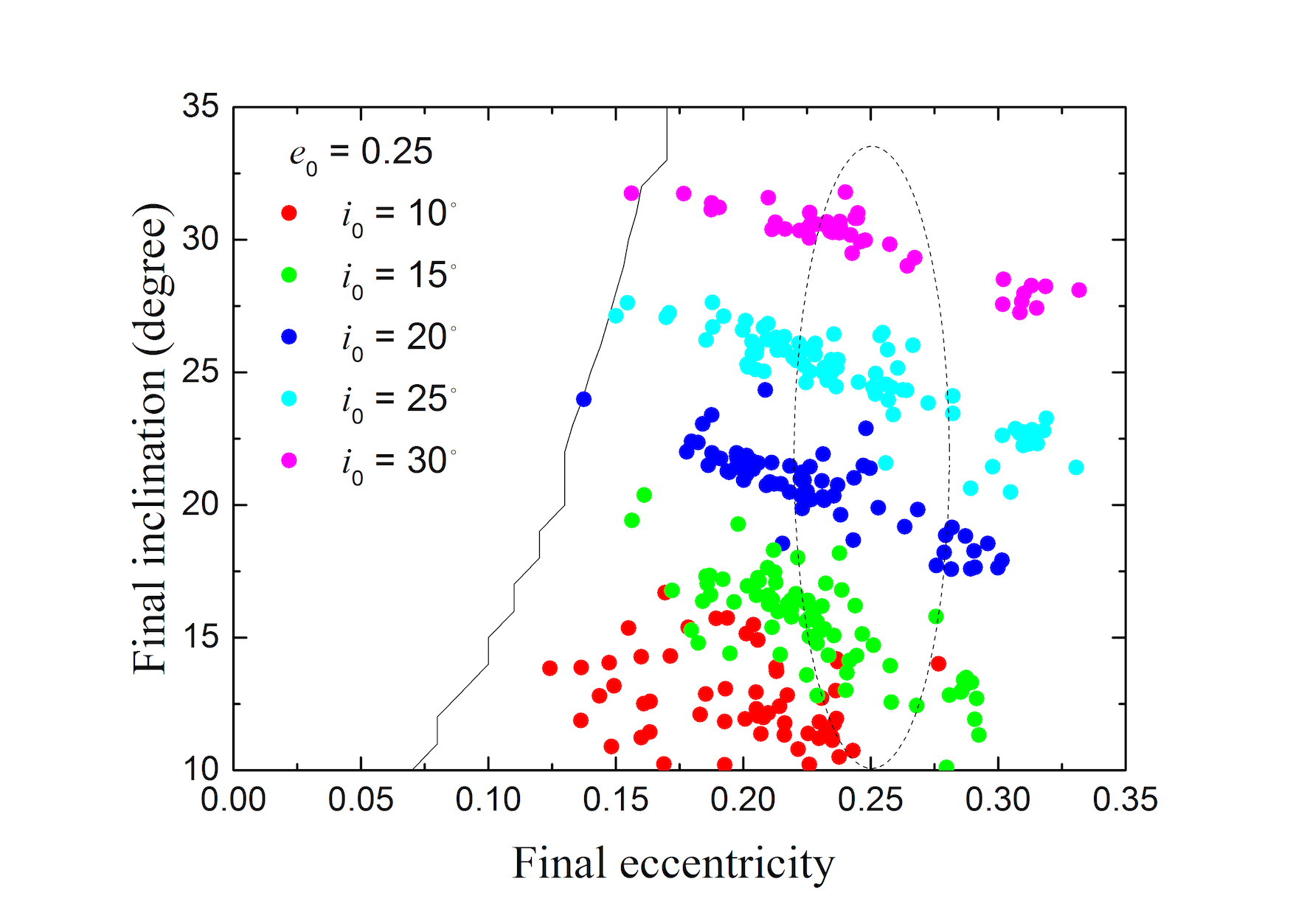}
  \end{minipage}
  \begin{minipage}[c]{1\textwidth}
  \vspace{0 cm}
  \includegraphics[width=9cm]{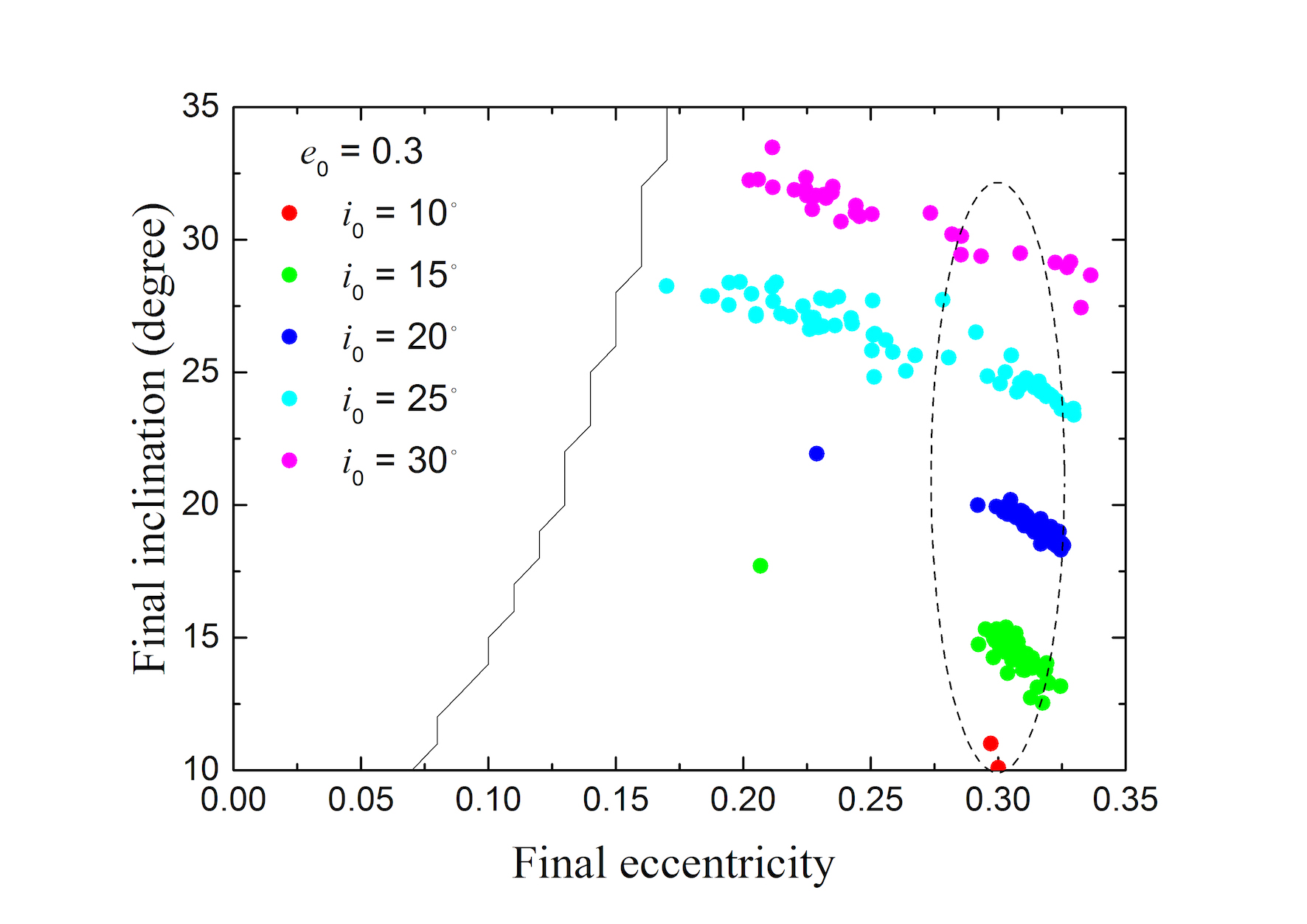}
  \end{minipage}
   \vspace{0 cm}
  \caption{Final eccentricities and inclinations of the resonant survivors in the 0-mode 4:7 MMR at the end of the 4 Gyr simulations, for initial eccentricities $e_0=0.1, 0.15, 0.2, 0.25$ and 0.3 (see labels in individual panels). In each panel, different colours indicate different initial inclinations $i_0$. Notice that nearly all the resonant survivors reside in the permissible region on the right side of the limiting curve (black line). In the $e_0=0.25$ panel, the blue dot lying on the limiting curve actually inhabits the permissible region, the apparent overlap is due to plot resolution. The other three points outside the permissible region are due to the Kozai mechanism, as discussed in the text. The dashed ovals indicate the resonant survivors that have final eccentricities close to their initial eccentricities.}
 \label{stability}
\end{figure*}

Numerical simulations are used to calculate the boundaries of the 0-mode 4:7 MMR, i.e., the maximum libration zones. These limits have been plotted on the initial $(a, i)$ plane in Fig.~\ref{boundary}(a). For a particular initial $e$, we see that the libration boundaries as a function of $a$ generally shrink with increasing $i$, and even disappear at high-$i$ where the particles are not librating at all. 
Taking the case of initial $e=0.1$ as an example, the corresponding resonance boundaries (green curves) are cut off at $i=15^{\circ}$. As illustrated in Fig.~\ref{boundary}(b), this is because librating particles (green dots) with higher $i$ can not be found above the limiting curve (black line). The numerical result appears perfectly consistent with the critical condition of $e=e_c^0(i=15^{\circ})=0.1$, indicated by the limiting curve.

In the case of initial $e=0.15$, the limiting curve predicts that the inclination of the resonant population could be up to $28^{\circ}$. This $i$-value neatly corresponds to the numerical separatrix layer between the librating particles with $i\le25^{\circ}$ (purple dots) and circulating ones with $i=30^{\circ}$ (purple cross) (see Fig.~\ref{boundary}(b)). With a higher resolution of $\Delta i=1^{\circ}$, a series of additional numerical calculations have been carried out in this specific $i$ range. We find that several $\sigma_0$-librators can be generated from the orbits with $i=27^{\circ}$, which is only $1^{\circ}$ smaller the theoretical upper limit, further supporting the validity of our limiting curve.

In the other three cases we explored for initial $e\ge0.2$, Fig.~\ref{boundary}(b) shows that the resonant particles (red, blue and orange dots) can be found at any $i$ up to $30^{\circ}$, which is the upper limit of the currently known 4:7 RKBOs. Since $e_c^0=0.2$ corresponds to a permissible $i$ much larger than $30^{\circ}$, the existence of the 0-mode resonators here is obvious and expected. It is noteworthy that, for high-$i$ particles, the resonant amplitudes may decrease to smaller values with increasing $e$ due to the less pronounced variation of the SLC. In this way, these eccentric particles with low perihelia could be better protected against strong gravitational perturbations from Neptune by the 4:7 MMR.

\subsection{The long-term evolution}

In this subsection, we further explore the long-term evolution of the high-inclination population in the 0-mode 4:7 MMR over 4 Gyr. The sets of initial eccentricity $e_0$ and inclination $i_0$ are adopted according to the librating particles in pre-runs: $e_0=0.1$ and $i_0=10^{\circ}-15^{\circ}$ ($\Delta i=1^{\circ}$);  $e_0=0.15$ and $i_0=10^{\circ}-25^{\circ}$ ($\Delta i=5^{\circ}$); $e_0=0.2, 0.25, 0.3$ and $i_0=10^{\circ}-30^{\circ}$ ($\Delta i=5^{\circ}$). For a specific pair of $(e_0, i_0)$, the initial semimajor axes $a_0$ are confined within the individual resonance boundaries shown in Fig.~\ref{boundary}(a). For the sake of saving computation time, we reduce the spatial resolution to $\Delta a_0=0.005$ au, which is still high enough to generate a number of resonant survivors for the analysis.

Our current objective is to comprehensively delineate the possible $(e, i)$ distribution of the potential 4:7 RKBOs. As librating particles from pre-runs may evolve on chaotic orbits during the long-term evolution, both their resonant behaviors and positions can change significantly. Thus the final orbital states of all particles are determined in the last 10 Myr of the integration, and the ``resonant survivors" are those particles with resonant amplitudes $A_{\sigma_0}<150^{\circ}$, which is approximately the upper limit for stable libration suggested by \citet{Lyka2005b}. Over the same time window, we calculate particles' average eccentricities and inclinations, which are recorded as the final parameters $e_f$ and $i_f$, respectively. Fig.~\ref{stability} summarizes the orbital distribution of the simulated resonant survivors with different initial conditions. In order to highlight the high-inclination population that is the topic of this paper, a few objects that end the simulation at $i_f<10^{\circ}$ have been discarded.

The overall result depicted in Fig.~\ref{stability} is that nearly all the resonant survivors still reside within the permissible region to the right of the limiting curve (black line), even after 4 Gyr of chaotic evolution. Out of hundreds of resonant survivors, there are only 3 that end up outside the permissible region due to the Kozai mechanism. This shows that our limiting curve can generally be used to constrain and predict the $(e, i)$ distribution for the 0-mode 4:7 MMR orbits. The potential 4:7 RKBOs at high inclinations may indeed cover the entire permissible region.
These long-term simulations show that chaotic evolution can drive the resonant survivors starting at the same $e_0$ onto orbits with quite a wide range of $e_f$, down to the lower limit $e_c^0(i_f)$ and up to $>0.3$. And, given a fixed $e_0$ in each panel, the objects with the same $i_0$ (i.e., the same colour) appear to have a visible trend of decreasing $i_f$ and increasing $e_f$. This distribution is mainly due to the coupling of $e$ and $i$ such that the quantity $K=\sqrt{1-e^2}\cos i$ (the z-component of angular momentum) could remain almost constant \citep{Li2014a}. According to our argument at the end of Section 3.1, such an orbital configuration of lower $i$ and larger $e$ benefits the long-term stability of the 0-mode resonators.

For the case of relatively small $e_0=0.1$, many resonant survivors have been excited to more eccentric orbits with $e_f\gtrsim0.15$ (top left panel in Fig.~\ref{stability}). Since their evolutionary paths are very complex, we will not examine each one in detail but concentrate on the final states. Most objects with $i_f$ between $10^{\circ}$ and $15^{\circ}$ only need an eccentricity of $e_f\ge0.1$ to guarantee the libration of the SLC. These resonators are always permitted and should be stable. However, among the resonant survivors on high-inclination orbits, very few of them are found to stay in the region of $e_f=0.1\pm0.025$, even including the contributions from the other cases of larger $e_0$. This suggests that the fraction of the resonant population with $i>10^{\circ}$ may be very low at $e\sim0.1$.

The case of moderate $e_0=0.15$ is shown in the upper right panel of Fig.~\ref{stability}. A few resonant survivors are not very mobile in eccentricity, remaining close to the original region near $e_f=0.15\pm0.025$. Among these, the typical value of $i_f$ is below $20^{\circ}$; while the most eccentric members with $e_f\sim0.175$ can achieve higher $i_f$ up to $25^{\circ}$. Nevertheless, a large fraction of the rest are found on the orbits with $e_f>0.2$, at the cost of decreasing inclinations by a few degrees. This is principally due to the (transitory) Kozai mechanism. 

\begin{figure}
 \hspace{-0.7cm}
 \vspace{-0.1cm}
  \centering
  \includegraphics[width=9cm]{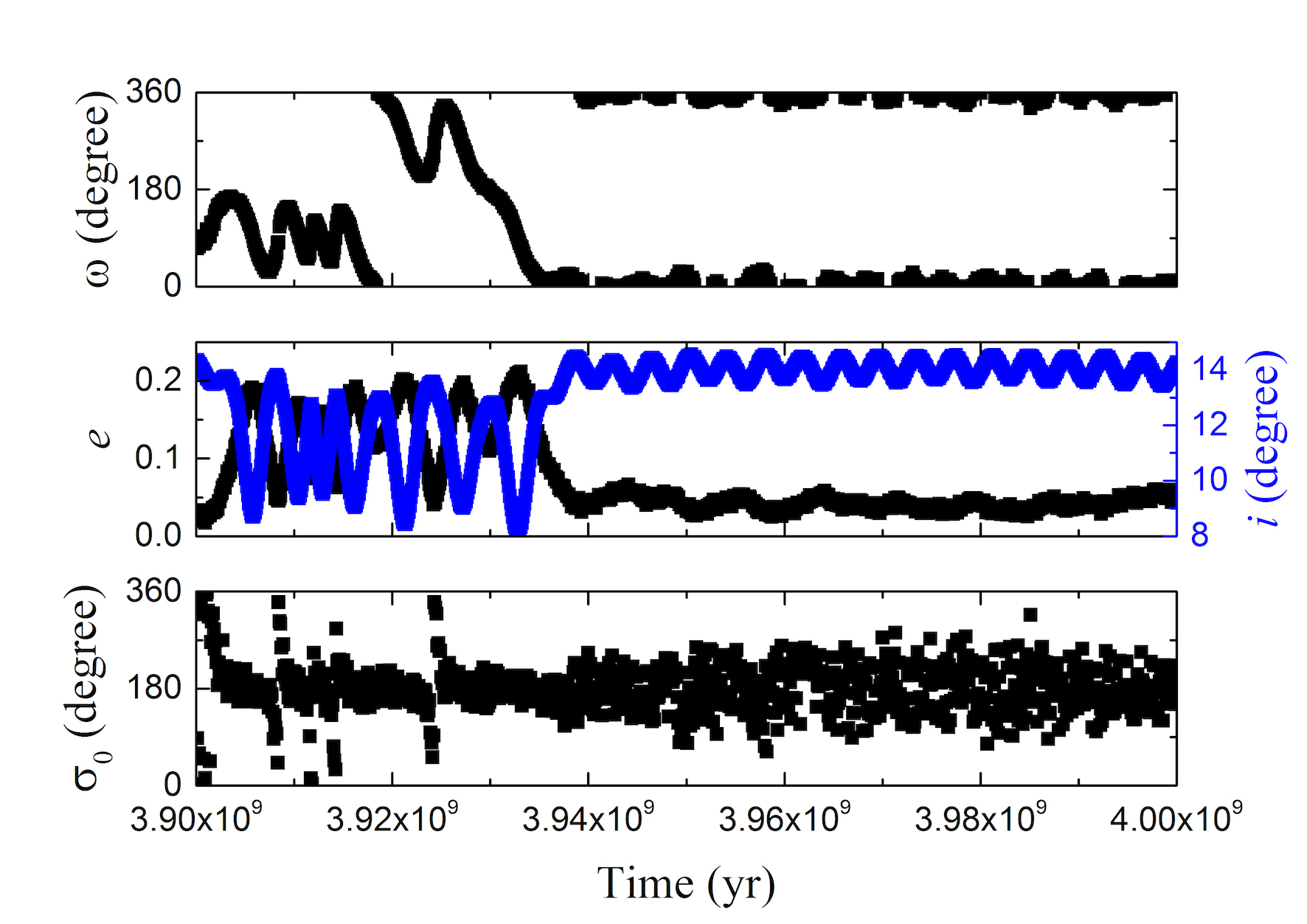}
  \caption{Example of a 0-mode 4:7 resonator that ends the simulation with $e_f$ and $i_f$ outside the permissible region. This exception is due to the libration of $\omega$ around $0^{\circ}$, i.e., the $K_{0}$-type Kozai mechanism, causing the SLC to be deeply locked near $180^{\circ}$ with very small amplitude even though the resonant condition $e\ge e_c^0$ is not fulfilled.}
  \label{exception}
\end{figure}

The Kozai mechanism appears to be important for inducing an eccentricity decrease and an inclination increase in some particles' orbits. In the simulation for $e_0=0.15$ with $i_0=10^{\circ}$, we observe that one of the resonant survivors settles to a final orbit with $e_f\approx0.05$ and $i_f=14^{\circ}$. In the upper right panel of Fig.~\ref{stability}, this exception is clearly visible as the red dot to the left of the limiting curve, i.e., outside the permissible region. It crossed the limiting curve in the last hundred million years of the integration, as displayed in Fig.~\ref{exception}. Between $3.90\times10^9$ and $3.92\times10^9$ yr, the particle experiences the $K_{90}$-type Kozai mechanism diagnosed by the libration of $\omega$ around $90^{\circ}$. During the same period, the eccentricity and inclination are both oscillating with very large amplitudes.
As a consequence, a low-$e$ state ($\sim 0.05<e_c^0(i)$) is achieved and the resonant amplitude $A_{\sigma_0}$ reaches $180^{\circ}$ at that time. But starting from about $3.94\times10^9$ yr, the particle re-encounters the Kozai mechanism and keeps its $\omega$ stably librating around $0^{\circ}$ until the end of the integration, with $e_f=0.05\pm0.02$ and $i_f=14^{\circ}\pm1^{\circ}$. At this final stage, the particle certainly violates the constraint of the limiting curve since $e_f< e_c^0(i=14^{\circ})=0.1$, but its resonant angle $\sigma_0$ can persistently librate. Just as we hypothesized at the end of Section 2.1, this simulation shows that such kind of exception is possible in the 0-mode 4:7 MMR due to the $K_{0}$- or $K_{180}$-type Kozai mechanism. Nevertheless, the odds of this occurring are very low, with only one example out of the 92 resonant survivors in the case of $e_0=0.15$.

At large $e_0$ $(\sim0.2-0.3)$, the inclination evolution of the resonant survivors seems quite regular (Fig.~\ref{stability}, lower three panels). A number of objects remain close to the region of $e_0$ and $i_0$ and still resemble their initial orbits at the end of the simulation. As a result of limited ($e$,$i$) mobility, these simulations should show the approximate ($e$,$i$) values of the potential 4:7 RKBOs. 
We thus propose that the probability of discovering the high-inclination 4:7 RKBOs should be greatest in the eccentricity interval of $[0.2, 0.3]$. 

In each case for the $e_0=0.2$ and 0.25 simulations, there is a single resonant survivor with $i_0=30^{\circ}$ (magenta dot) outside the permission region, with the total number of resonant survivors being 215 and 330, respectively. 
This further suggests that the probability of leaving the permissible region is extremely low. 
We confirm that both of these two exceptions experience libration of $\omega$ around $0^{\circ}$, the same as in the case of $e_0=0.15$ (Fig.~\ref{exception}). So to refine our prediction from earlier based on these numerical results, the limiting curve appears not to work well for 4:7 resonators experiencing the $K_{0}$-type (or probably $K_{180}$-type) Kozai mechanism.  
Alternatively, the limiting curve is very good at predicting that any 4:7 resonators outside the permissible region are associated with the Kozai mechanism.

\subsection{Summary of dynamical evolution results} \label{sec:dynsum}

In Section~\ref{sec:dyn}, we performed numerical simulations to measure the limits of the 0-mode 4:7 MMR in $(a,e,i)$-space, and explore the dynamical evolution of simulated particles over 4 Gyr. 
As expected, we found that the libration width is narrower in $a$ at low-$e$ and high-$i$, and the limiting curve $e^0_c(i)$ prediction holds as particles having $e=0.1$ with $i>15^{\circ}$, and $e=0.15$ with $i>27^{\circ}$ do not librate in the 4:7 resonance.
Long term evolution of orbits within the resonance varies widely depending on initial eccentricity $e_0$ and inclination $i_0$. But generally, higher $e_0$ and $i_0$ particles are less mobile in ($e$,$i$)-space and stick close to their initial orbits for the age of the Solar system. Taken in total, hundreds of simulated 0-mode 4:7 resonators mostly remain in the permissible region lying to the right of the limiting curve. Although a few resonators enter the non-permissible region due to the Kozai mechanism, we predict this is not a very important aspect in the potential 4:7 RKBOs distribution since the probability is very low in our simulations ($<$1\%).


\section{Resonance capture}

Following \citet{Li2014a, Li2014b}, we now conduct numerical experiments on the 4:7 MMR capture of test particles in the framework
of the planet migration scenario \citep[e.g.][]{Fern84, Malh1995}. Our goals here are to further investigate the following questions: \begin{enumerate}
\item Can the particles captured into the 0-mode resonance fill the initial eccentricity and inclination space $(e_0, i_0)$ found for the long-term evolution analysis above? 
\item For the 0-mode resonance, how do the initial $i$-values of test particles affect the probability of resonant capture and retention, and the occurrence of the Kozai mechanism? 
\item  \citet{Morb2014} showed that, with $e=0.1$ and $i=19^{\circ}$, particles cannot be captured into the 4:7 MMR. Is this outcome related to the fact that, at such a high $i$, this  small $e$ is below the critical eccentricity (i.e.,$< e_c^0(i=19^{\circ})=0.12$), which would prohibit the libration of $\sigma_0$?
\item Can any particles be independently captured into the $-1$-mode of the 4:7 MMR?
\end{enumerate}

In order to highlight the capture of test particles into the 4:7 MMR, inspired by \citet{Morb2014}, we consider the end of the planet migration phase: Jupiter, Saturn and Uranus had reached their current orbits; while Neptune was on the nearly circular and co-planar orbit at $a_i^N=28.2$ au, and moved smoothly outward to its current location of $a_f^N=30.2$ au according to
\begin{equation}
  a^N(t)=a_f^N-\Delta a^N \exp(-t/\tau),
\label{neptune}
\end{equation}
where the net radial displacement is $\Delta a^N=a_f^N-a_i^N=2$~au, much smaller than the value of $\sim7$ au used in \citet{Li2014a, Li2014b}. Under this set-up, given the same migration timescale of $\tau=2\times10^7$ yr, the migration speed of Neptune is about 3.5 times slower. Here we use a smaller, slower migration in order to enhance the capture probability of the weak 4:7 MMR, allowing enough particles to be captured into this resonance for the statistical analysis below. During the simulated migration, Neptune's 4:7 MMR will move from an initial position of $28.2\times(7/4)^{2/3}=40.9$~au to its current position of 43.8~au. 

To efficiently distribute test particles in the path of this sweeping resonance's migration, we spread all particles in the region of $a=41.9-44$ au: from 1 au exterior to the initial location of the 4:7 MMR out to just a little beyond the current 4:7 MMR location. This range of initial $a$-values is within the area of the MCKB, so these simulations also simultaneously test resonance sweeping through the primordial planetesimal disk. Due to computational limits, we did not cover the entire range of possible initial eccentricities. Instead, with particular focus on the limiting curve for the 0-mode 4:7 MMR, we designed a series of simulations comprised of test particles with representative initial eccentricities\footnote{In order to distinguish from the initial (final) inclinations and eccentricities $i_0$ ($i_f$) and $e_0$ ($e_f$) in Section 3, in the discussion of the migration model we will use $I_0$ ($I_f$) and $E_0$ ($E_f$) instead.} $E_0$ = (0.05, 0.1, 0.15, 0.2, 0.25). 

For each given $E_0$, we performed 4 identical runs with initial inclinations $I_0=(1^{\circ}, 10^{\circ}, 20^{\circ}$, $30^{\circ}$). These runs have the same number of test particles (417) with $\Delta a=0.005$ au in order to cover the $a$ range. We integrated each system for $10^8$ yr, which is long enough for Neptune to reach its current location and orbit there without migration for several tens of Myr. Particles are removed from the simulations if their heliocentric distances pass inside the orbit of Jupiter or exceed 100 au. For surviving particles, we measured both 4:7 resonant angles ($\sigma_0$ and $\sigma_{-1}$) over the last 10 Myr, and those objects with resonant amplitudes smaller than $150^{\circ}$ are regarded as ``captured 4:7 resonators''. Their reported final eccentricities $E_f$ and inclinations $I_f$ are also averaged over the last 10~Myr. Below, we will separately present the results for captured 4:7 resonators in the 0-mode (Section~\ref{sec:0mode}) and $-1$-mode (Section~\ref{sec:m1mode}) of the 4:7 MMR. 

\subsection{Results for the 0-mode 4:7 MMR} \label{sec:0mode}

\begin{figure}
 \hspace{-0.7cm}
  \centering
  \includegraphics[width=9cm]{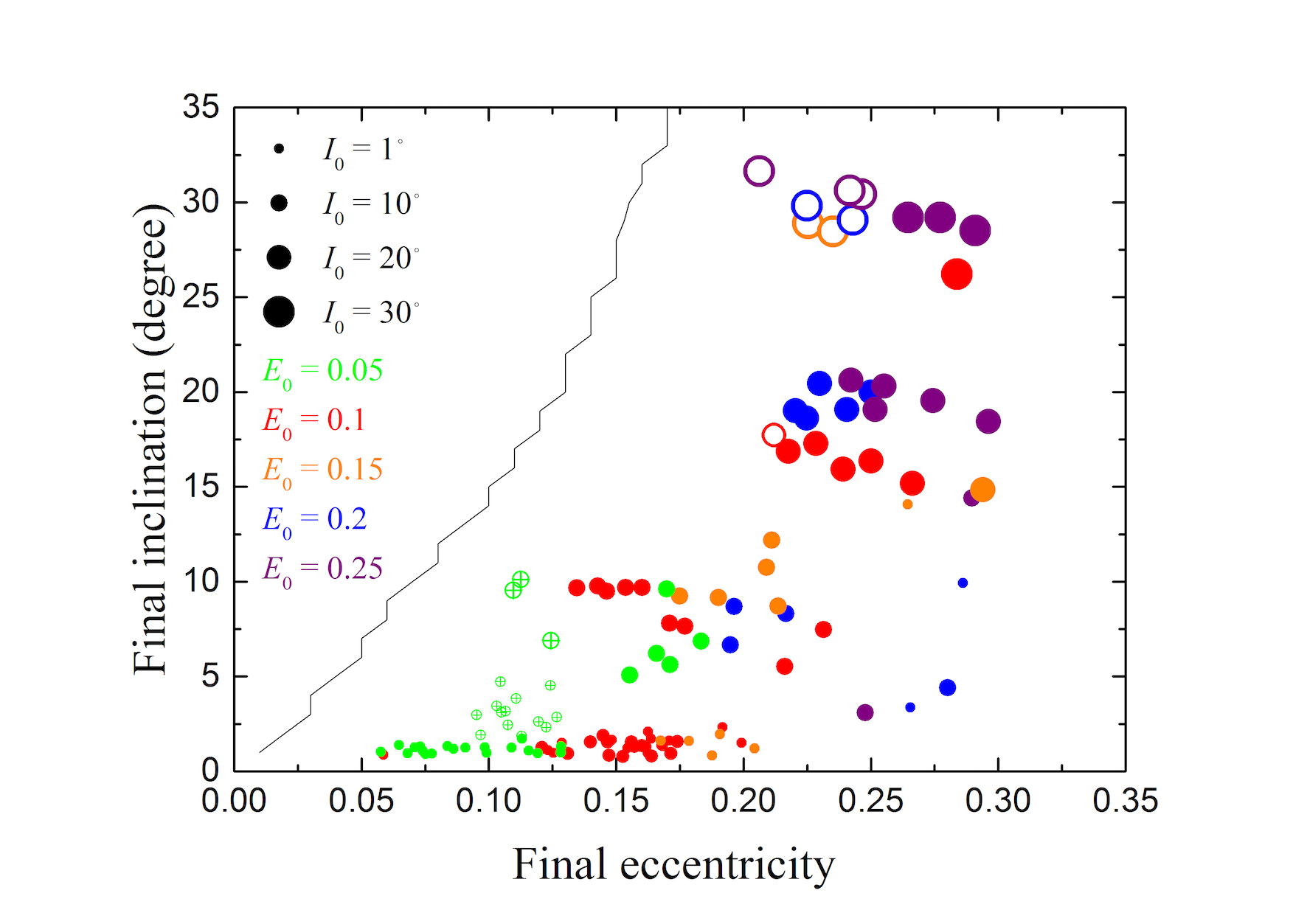}
  \caption{Final eccentricities and inclinations of captured 0-mode 4:7 resonators within the planet migration model. The initial eccentricities $E_0$ of test particles are indicated by different colours. At each $E_0$, we adopt several initial inclinations $I_0$ from 1$^{\circ}$ to 30$^{\circ}$, which are depicted by different sizes of the dots. Open dots indicate $K_{0}$- or $K_{180}$-type Kozai resonators, and open dots with plus symbols inside indicate $K_{90}$- or $K_{270}$-type ones. As we expected, all the captured resonators are found well inside the permissible region, i.e., on the right side of the (black) limiting curve.}
  \label{capture0}
\end{figure}

Fig.~\ref{capture0} summarizes the final eccentricities and inclinations of captured resonators in the 0-mode 4:7 MMR for different pairs of initial $(E_0, I_0)$. It can be immediately observed that, within the planet migration model, all these captured resonators fall into the permissible region on the right side of the limiting curve (black line). Even objects that started with $E_0\ll e_c^0(I_0)$ had their eccentricities pumped up and inclinations held nearly constant while they were forced outwards in the migrating resonance. At the end of the simulation, they meet the requirement of $E_f\ge e_c^0(I_f)$.

Captured resonators with final eccentricities $E_f\lesssim0.1$, coming from the $E_0=0.05$ and 0.1 cases (green and red dots in Fig.~\ref{capture0}), could have final inclinations $I_f$ up to $10^{\circ}$. According to the long-term simulations in Section 3, some of these objects may remain around $e\sim0.1$ and later evolve to higher $i\sim15^{\circ}$. Even as they evolve, the $(e, i)$ distribution originating from the $E_f\lesssim0.1$ population continues to be constrained by the limiting curve.

The captured resonators with larger $E_f$ ($\sim0.2-0.3$) cover the whole high-inclination range of $10^{\circ}\le I_f\le30^{\circ}$. The simulations in Section 3 show that these highly eccentric objects can keep their $i$ mostly unchanged over the age of the Solar system, while their $e$ may disperse to a wider range of $0.15-0.3$. This implies that the deficit of the 4:7 resonators in the region of $I_f>10^{\circ}$ at $E_f\sim0.15$ will eventually be compensated by the slow orbital diffusion.

\begin{table}
\centering
\begin{minipage}{8.5cm}
\caption{The statistics of captured resonators for initial inclinations $I_0$ from 1$^{\circ}$ to 30$^{\circ}$ within the planet migration model, in terms of numbers of objects. Several representative values of initial eccentricities $E_0$ are adopted. For each pair of $(E_0, I_0)$ in a run, there are 417 test particles with initial semimajor axes distributed uniformly between 41.9 and 44 au. The second column ($N_{4:7}^0$) refers to particles captured into the 0-mode 4:7 MMR; among them, objects experiencing the Kozai mechanism with librations of $\omega$ around $0^{\circ}$ or 180$^{\circ}$ ($K_{0\&180}^0$) and 90$^{\circ}$ or 270$^{\circ}$ ($K_{90\&270}^0$) are indicated in the third and fourth columns, respectively. The fifth column ($N_{4:7}^{-1}$) refers to particles trapped into the $-1$-mode 4:7 MMR. The last column refers to the Kozai objects associated with the -1-mode, only $K_{0}$- and $K_{180}$-type resonators ($K_{0\&180}^{-1}$) are found.}      
\label{resonators}
\begin{tabular}{c c c c c c c}        
$E_0=0.05$      \\
\hline\hline
$I_0(^{\circ})$       &         $N_{4:7}^0$       &   $K_{0\&180}^0$      &  $K_{90\&270}^0$       &       $N_{4:7}^{-1}$    &   $K_{0\&180}^{-1}$           \\
\hline
1                                &               32                   &                0                &                13                &                   0                            &               0                                 \\
10                              &               8                     &                0                &                 3                 &                   1                            &               0                                 \\
20                              &               0                     &                0                &                 0                 &                   1                            &               0                                 \\
30                              &               0                     &                0                &                 0                 &                   5                            &               0                                 \\
\hline\hline\\
$E_0=0.1$      \\
\hline\hline
$I_0(^{\circ})$       &         $N_{4:7}^0$       &   $K_{0\&180}^0$      &  $K_{90\&270}^0$       &       $N_{4:7}^{-1}$     &   $K_{0\&180}^{-1}$          \\
\hline
1                                &               29                   &                0                &                  0                &                   0                             &               0                                 \\
10                              &               9                     &                0                &                  0                &                   0                             &               0                                 \\
20                              &               6                     &                1                &                  0                &                   3                             &               1                                 \\
30                              &               1                     &                0                &                  0                &                   5                             &               0                                 \\
\hline\hline\\
$E_0=0.15$      \\
\hline\hline
$I_0(^{\circ})$       &         $N_{4:7}^0$      &   $K_{0\&180}^0$      &  $K_{90\&270}^0$       &       $N_{4:7}^{-1}$     &   $K_{0\&180}^{-1}$          \\
\hline
1                                &                 6                   &                0                &                  0                &                   0                             &               0                                 \\
10                              &                 5                   &                0                &                  0                &                   0                             &               0                                 \\
20                              &                 1                   &                0                &                  0                &                   2                             &               1                                 \\
30                              &                 2                   &                2                &                  0                &                   1                             &               0                                 \\
\hline\hline\\
$E_0=0.2$      \\
\hline\hline
$I_0(^{\circ})$       &         $N_{4:7}^0$       &   $K_{0\&180}^0$      &  $K_{90\&270}^0$       &       $N_{4:7}^{-1}$      &   $K_{0\&180}^{-1}$        \\
\hline
1                                &                 2                   &                0                &                  0                &                   0                              &               0                                 \\
10                              &                 4                   &                0                &                  0                &                   0                              &               0                                 \\
20                              &                 5                   &                0                &                  0                &                   1                              &               0                                 \\
30                              &                 2                   &                2                &                  0                &                   0                              &               0                                 \\
\hline\hline\\
$E_0=0.25$      \\
\hline\hline
$I_0(^{\circ})$       &         $N_{4:7}^0$       &   $K_{0\&180}^0$      &  $K_{90\&270}^0$       &       $N_{4:7}^{-1}$       &   $K_{0\&180}^{-1}$     \\
\hline
1                                &                 0                   &                0                &                  0                &                   0                                &               0                                 \\
10                              &                 2                   &                0                &                  0                &                   0                                &               0                                 \\
20                              &                 5                   &                0                &                  0                &                   2                                &               0                                 \\
30                              &                 6                   &                3                &                  0                &                   1                                &               0                                 \\
\hline\hline\\
\end{tabular}
\end{minipage}
\end{table}

This first glance at the results of migration simulations positively answers question (i) at the beginning of this section: it appears that captured resonators from a small range of ($E_0$,$I_0$) can indeed fill the initial ($e_0$, $i_0$) space in the permissible region. We propose that a future survey would likely discover 0-mode 4:7 RKBOs anywhere on the right side of the limiting curve $e_c^0(i)$. Currently, very few 4:7 RKBOs have been detected in the high-$i$ region, as shown in Fig.~\ref{PR}(a). Linked to question (ii), below we discuss in detail some trends observed in the 0-mode captured resonators for different values of $I_0$, and the associated Kozai mechanism.

\subsubsection{General trends in 0-mode capture}

Details of the dependence of resonant capture on particles' initial inclinations $I_0$ are provided in Table~\ref{resonators}:\\

(1) \underline{At small $I_0=1^{\circ}$:} the number of captured resonators ($N_{4:7}^0$) falls from 32 for $E_0=0.05$ to 0 for $E_0=0.25$. This $e$-dependence is consistent with the adiabatic invariant theory described in \citet{Meli2000}. For an exterior MMR, the capture probability is 100\% when test particles approach the resonance with $e$ smaller than a critical value of $e_{\mbox{\footnotesize{crit}}}$, and it drops very rapidly when $e>e_{\mbox{\footnotesize{crit}}}$. For the 1:2, 2:3 and 3:5 MMRs, the values of $e_{\mbox{\footnotesize{crit}}}$ were determined to be 0.061, 0.052, and 0.033, respectively: apparently decreasing for higher-order resonances. Thus we expect that the 4:7 MMR should have a value of $e_{\mbox{\footnotesize{crit}}}^{4:7}$ even smaller than 0.033. This concept is useful to explain the low capture rate for the low-inclination population with $E_0\ge0.05$($>e_{\mbox{\footnotesize{crit}}}^{4:7}$). However, the exact $e_{\mbox{\footnotesize{crit}}}^{4:7}$ depends on the migration speed of Neptune, and we will not try to calculate it as this goes beyond the scope of the present work.

(2) \underline{At moderate $I_0\sim10^{\circ}-20^{\circ}$}: For such higher $I_0$, the capture efficiency of the 0-mode 4:7 MMR (denoted by the number $N_{4:7}^0$) shows no clear pattern as $E_0$ increases. We think this could be due to the competition between two mechanisms: increased chance of resonance capture at lower $e$ (as discussed above), and larger $e$ resonators being more sustainable to be inside the permissible region for a given $i$.

Taking the case of $I_0=20^{\circ}$ as an example, at $E_0=0.1$, we find that some test particles were captured by the migrating 4:7 MMR, but a large fraction of them quickly leaked out of this resonance. This leakage can be due to the fact that, at the very beginning of resonance capture, these inclined particles were outside the permissible region as they had $e\sim E_0 < e_c^0(i\sim I_0)=0.13$, which should cause their resonant angles $\sigma_0$ to circulate. Nevertheless, $N_{4:7}^0=6$ particles managed to sustain librating $\sigma_0$ while having $e$ excited in the migrating resonance. After migration, they successfully evolved to the stable libration states as the final eccentricities had entered the permissible region of $E_f\ge  e_c^0$. This example also points out that the initial condition $E_0 < e_c^0$ does not completely prohibit capture by the migrating 4:7 MMR. As envisioned in \citet{Morb2014}, for a high-order resonance having multiple modes, the temporary resonant capture of high-inclination particles may be still possible because of chaotic motion.

In the same case of $I_0=20^{\circ}$, fewer particles enter the 4:7 MMR at relatively larger $E_0=0.15$ simply because higher $e$ likely induces lower capture probability as mentioned before. So, the number of captured resonators with this ($E_0$,$I_0$) pair is only $N_{4:7}^0=1$. But if we continue to increase $E_0$ to $\ge0.2$, that is much larger than the critical eccentricity $e_c^0 (i=20^{\circ})$, and thus can significantly enhance the maintenance of resonant objects during the migration. As a result, the population of captured resonators $N_{4:7}^0$ $(= 5)$ becomes larger.

(3) \underline{At the largest $I_0=30^{\circ}$}: the number $N_{4:7}^0$ of captured resonators increases from 0 for $E_0=0.05$ to 6 for $E_0=0.25$. This trend is totally opposite to that found in the nearly co-planar case where $I_0=1^{\circ}$. Here the largest $E_0$ not only fulfills the condition $e\ge e_c^0(i)$ for libration of $\sigma_0$, but is also likely to result in quite small resonant amplitudes; thus captured resonators are more easily produced. Here, the mechanism based on the limiting curve appears to dominate the capture of 0-mode resonators by the 4:7 MMR.

(4) \underline{At high $E_0=0.25$}: For captured resonators, the number $N_{4:7}^0$ increases monotonically with $I_0$. Their $e$ values are much larger than the maximum $e_c^0(i=30^{\circ})=0.16$ during the entire evolution, so the limiting curve cannot explain this trend. The alternative explanation may lie in the minimum distance between a particle and Neptune, which can be estimated by
\begin{equation}
d=a(1-E_0)-a_N,
\label{CE}
\end{equation}
where $a_N=a \cdot\sqrt[3]{(4/7)^2}$ is Neptune's semimajor axis at the moment of the 4:7 MMR encounter. Given the initial $a$-range of 41.9-44~au, particles could have $d$ as small as 2.7 au, and thus suffer strong perturbations from Neptune. But the more inclined objects spend less time near Neptune's orbital plane, they are more likely to stay on nearly unperturbed orbits until the resonance passes.

These detailed capture simulations help provide an answer to question (iii) raised at the beginning of Section 4: \citet{Morb2014} showed that for two examples of particles with $E_0\sim0.1$, the low-inclination one ($I_0=0.5^{\circ}$) can be captured into the migrating 4:7 MMR, but the high-inclination one ($I_0=19^{\circ}$) cannot. We suspect that the failure to capture the latter may be ascribed to the limiting curve, which at this ($e, i$) pair corresponds to $E_0\sim0.1< e_c^0(I_0=19^{\circ})=0.12$. In our $E_0=0.1$ simulations, at initial inclinations $I_0=1^{\circ}$ and $20^{\circ}$ comparable to those of \citet{Morb2014}, the numbers $N_{4:7}^0$ of captured resonators are 29 and 6, respectively. This suggests that the capture probability for high-inclination particles is much lower, but it is still possible. We note that the simulations in \citet{Morb2014} use an initially eccentric Neptune, which had its eccentricity damped to $\sim 0.02$ when the 4:7 MMR passed their two particles. Hence, our simulation with a near-circular Neptune is valid for comparison.

Before finishing this subsection, we must highlight that the captured resonators in our simulations get more and more scarce for higher values of $E_0$, so the statistics could be somewhat limited because of the small number of samples. Nevertheless, our results broadly show the importance of a particle's initial inclination $I_0$ on the 0-mode 4:7 MMR capture.

\subsubsection{Kozai mechanism}

Many of the captured resonators in our simulations temporarily experienced the Kozai mechanism after entering the 0-mode 4:7 MMR. This dynamical phenomenon can cause a large increase in eccentricity, which may bring particles deeper inside this resonance as a result of smaller amplitudes of the SLCs. But we found that only a very small number of captured 4:7 resonators still have their $\omega$ librating at the final 10~Myr of the integration, and they are recorded in Table~\ref{resonators}. In the following we will explore the plausible distribution of these $\omega$-librators in detail.

We first consider the captured resonators that have final $\omega$ librating around $0^{\circ}$ or $180^{\circ}$: the $K_{0}$- and $K_{180}$-type Kozai resonators. Their numbers are indicated by $K_{0\&180}^0$ in Table~\ref{resonators}. There are a total of 8 such $\omega$-librators found in our simulations, and they all originated from high-$i$ orbits. According to the discussion in Section 2.1, these $K_{0}$ and $K_{180}$-type $\omega$-librators may be restricted to very small variations of the SLCs, and possibly present themselves as the exceptions which reside in the 0-mode 4:7 MMR but are on the wrong side of the limiting curve, with $e<e_c^0$. While in our simulations, none are observed outside the permissible region (open circles in Fig.~\ref{capture0}), in agreement with the real 4:7 RKBOs (open circles in Fig.~\ref{PR}(a)). This could be attributed to the dynamical process that, during outward migrations driven by the 4:7 MMR, initially captured particles with $e<e_c^0$ may have already left this resonance before the Kozai mechanism can occur. In the low-inclination regime of $i<10^{\circ}$, we observe a significant deficit of the $K_{0}$- or $K_{180}$-type  $\omega$-librators in the migration simulation. Even though large mobility in $(e, i)$ space is predicted due to the later chaotic diffusion in the 4:7 MMR \citep{Lyka2005b}, low-$i$ Kozai resonators evolving from the high-$i$ regime ($\sim20^{\circ}-30^{\circ}$) seems quite unlikely. Future more detailed work focusing on the low-inclination population may tell.

Next we look at the captured resonators that have $\omega$ librating around $90^{\circ}$ or $270^{\circ}$: the $K_{90}$- and $K_{270}$-type Kozai resonators. In total, we find 16 of these particles, indicated by the number $K_{90\&270}^0$ in Table~\ref{resonators}. As discussed in Section~2.1, the 4:7 MMR condition $e\ge e_c^0$ has to be met for $K_{90}$- and $K_{270}$-type particles, otherwise the small amplitude $\Delta\omega$ cannot prevent the circulation of the SLC. Therefore, all of the captured resonators experiencing the $K_{90}$- or $K_{270}$-type Kozai mechanism from the migration simulation are found in the permissible region (Fig.~\ref{capture0}, circles with plus symbols inside). Moreover, we notice that these $\omega$-librators in general have $I_f\lesssim10^{\circ}$, consistent with the observed samples (Fig.~\ref{PR}(a), circles with plus symbols inside).

The Kozai mechanism typically takes place at high inclination, and there is a well-known low limit of $i=\arccos({\sqrt{3/5})}\approx39.2^{\circ}$ for $e=0$ \citep{Kino2007}. However, inside the 2:3 MMR, the Kozai critical inclination can be as small as $\sim10^{\circ}$ \citep{Li2014a}.
A striking feature of the Kozai mechanism inside the 4:7 MMR is that 4 out of the 24 $\omega$-librators from our simulations emerge at extremely low inclinations: $i\lesssim2^{\circ}$. This is also true for the observed Kozai 4:7 RKBOs, as 3 out of 13 are found at $i\lesssim2^{\circ}$. We thus propose that the Kozai mechanism can operate inside the 4:7 MMR at any inclination, as long as the eccentricity is not too small (i.e., $e\gtrsim0.1$).

\subsection{Results for the $-1$-mode 4:7 MMR} \label{sec:m1mode}

We here discuss the captured resonators associated with the $-1$-mode of the 4:7 MMR in our migration simulations, and their numbers are recorded as $N_{4:7}^{-1}$ in Table~\ref{resonators}.
These objects are found to survive with $\sigma_{-1}$ librating around $180^{\circ}$ in a stable state for the last 10~Myr of the integration, while the other resonant angle $\sigma_0$ is circulating. As in Section~2.2, this configuration is defined as the independent $-1$-mode resonance. 
These simulations provide the answer to question (iv) posed at the beginning of Section 4: we find that test particles can indeed be captured into the 4:7 resonance and be independently librating in the $-1$-mode.

\subsubsection{General trends in $-1$-mode capture}

At $I_0=1^{\circ}$, none of test particles can be trapped into this resonance (i.e., $N_{4:7}^{-1}=0$ in Table~\ref{resonators}). A straightforward interpretation is that the resonance's strength is rather weak for low-inclination orbits, because the coefficient of the resonant term follows the function $ei^2$. Thus, for high-inclination orbits, the strength of the $-1$-mode resonance would grow considerably and capture should be more likely. 

As shown in Table~\ref{resonators}, at $I_0\ge10^{\circ}$,  capture into the $-1$-mode resonance does appear to always be possible. Adding together all the cases of $E_0\ge0.05$, the number ratio $N_{4:7}^{-1}$/$N_{4:7}^{0}$ between the $-1$-mode and 0-mode captured resonators is about 0.17. This value seems quite large compared to observations, since only a single $-1$-mode member has been identified so far out of the 34 known 4:7 RKBOs. Although the currently observed ratio of $N_{4:7}^{-1}$/$N_{4:7}^{0}$ is biased due to incomplete observations, the $-1$-mode population is probably still overrepresented in our simulations. This issue may be due to the initial eccentricities and inclinations of test particles that we adopted, and the ratio $N_{4:7}^{-1}$/$N_{4:7}^{0}$ could actually provide constraints on future migration modeling. 

Table~\ref{resonators} shows that, the simulated $-1$-mode resonators mostly originate from less eccentric ($E_0=0.05$ and 0.1) but more inclined ($I_0=20^{\circ}$ and $30^{\circ}$) orbits. In fact, all of these objects started within the $-1$-mode permissible region of $e\le e_c^{-1}(i=20^{\circ})=0.12$, which made them more likely to be captured. Interestingly, their 0-mode counterparts are initially outside the corresponding permissible region of $e\ge e_c^{0}(i=20^{\circ})=0.13$, where the capture efficiency should not be high. This difference in starting conditions relative to the limiting curve could be what leads to the relatively large value of $N_{4:7}^{-1}$/$N_{4:7}^{0}$ in our simulations. 

\begin{figure}
 \hspace{-0.7cm}
  \centering
  \includegraphics[width=9cm]{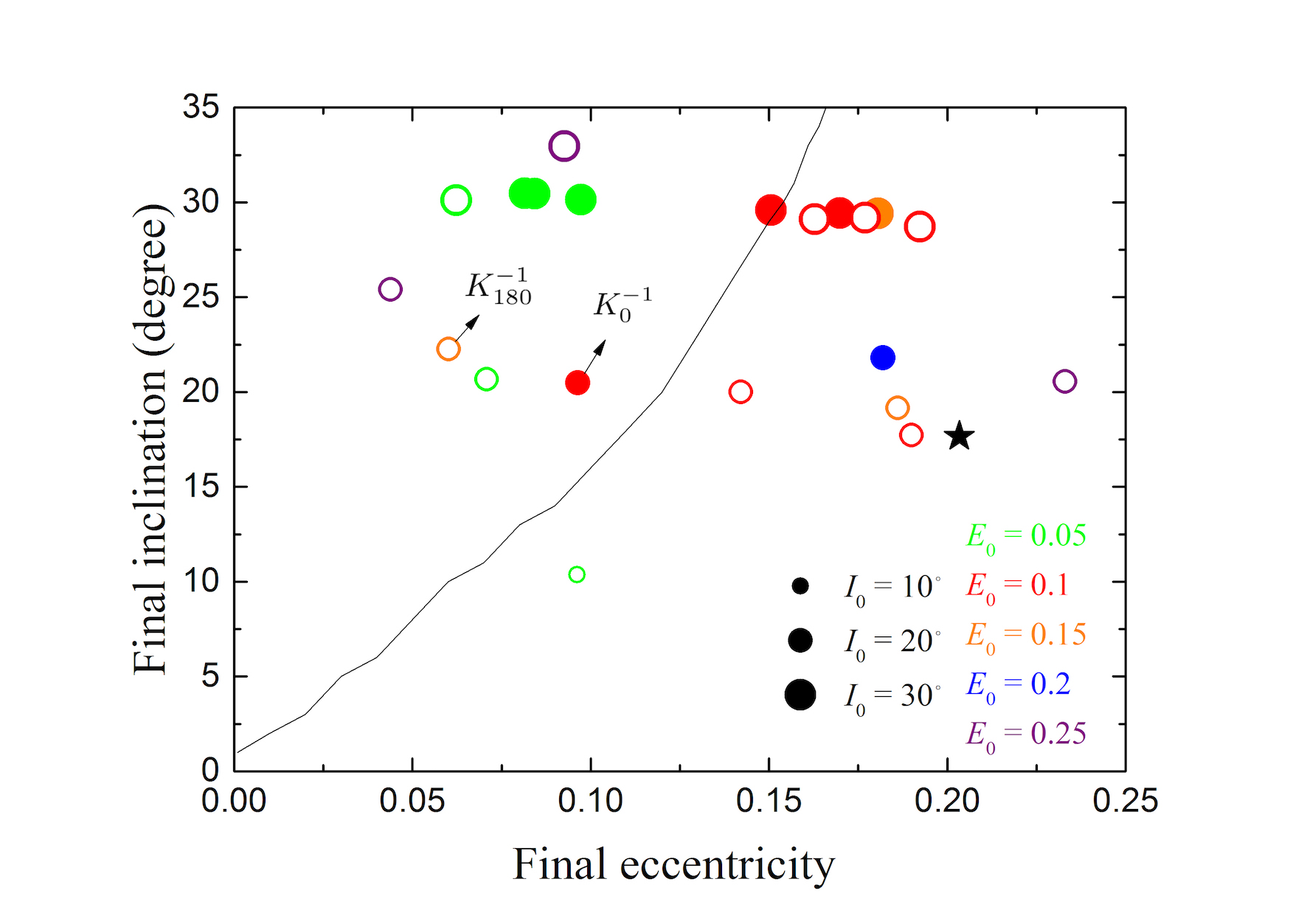}
  \caption{Similar to Fig.~\ref{capture0}, but for captured resonators associated with the $-1$-mode 4:7 MMR. The filled circles indicate the objects that can remain in this resonance in the extended integrations up to 1 Gyr, while open circles indicate those that can not. Note that the $-1$-mode captured resonators are supposed to reside to the left of the (black) limiting curve, to sustain the libration of $\sigma_{-1}$. But obviously there are considerable exceptions, including the real RKBO 2014 TZ85 represented by the star. The two objects experiencing the Kozai mechanism with librations of $\omega$ around $0^{\circ}$ and $180^{\circ}$ are indicated by $K_{0}^{-1}$ and $K_{180}^{-1}$, respectively.}
  \label{capture-1}
\end{figure}

Fig.~\ref{capture-1} displays the final eccentricities and inclinations of the $-1$-mode captured resonators. It is worth mentioning that their inclinations evolve with only minor changes during the integration, i.e., keeping $I_f\sim I_0$. The most populated region seems to be at $i\gtrsim20^{\circ}$, where the real observed $-1$-mode RKBO 2014 TZ85 resides (indicated by the star). 

In our simulations, we observe that there is only one $-1$-mode resonator with inclination as low as $i\sim10^{\circ}$. This object had its resonant angle $\sigma_{-1}$ transition from circulation to libration about 10 Myr before the end of the integration, which is the same timespan we used to identify the resonant behavior. Was it periodically trapped into the $-1$-mode resonance, or just a transient resident? What about the long-term stability of the other higher-$i$ resonators? To better explore their resonant behaviors, we extended the integration of these captured resonators up to 1 Gyr. We find that the evolution of the lowest-$i$ ($\sim10^{\circ}$) simulated resonator is extremely chaotic, and it left the $-1$-mode resonance forever in less than 1 Myr after starting the extended run.
We argue that the instability is because the strength of this mixed-$(e, i)$-type resonance is not strong enough at a relatively low $i$-value. The particles with $i\gtrsim20^{\circ}$ remain in the $-1$-mode resonance at least for another 50 Myr, and about 40$\%$ can keep their $\sigma_{-1}$ persistently librating until 1 Gyr (indicated by filled circles in Fig.~\ref{capture-1}). Thus we predict that future surveys focused on discovering KBOs at high ecliptic latitudes will find more 4:7 RKBOs belonging to the $-1$-mode in addition to 2014 TZ85. The updated total number of $-1$-mode 4:7 RKBOs could provide very useful constraints on the planetesimal disk that the 4:7 MMR swept through, including the original MCKB.

\begin{figure}
 \hspace{-0.7cm}
  \centering
  \includegraphics[width=9cm]{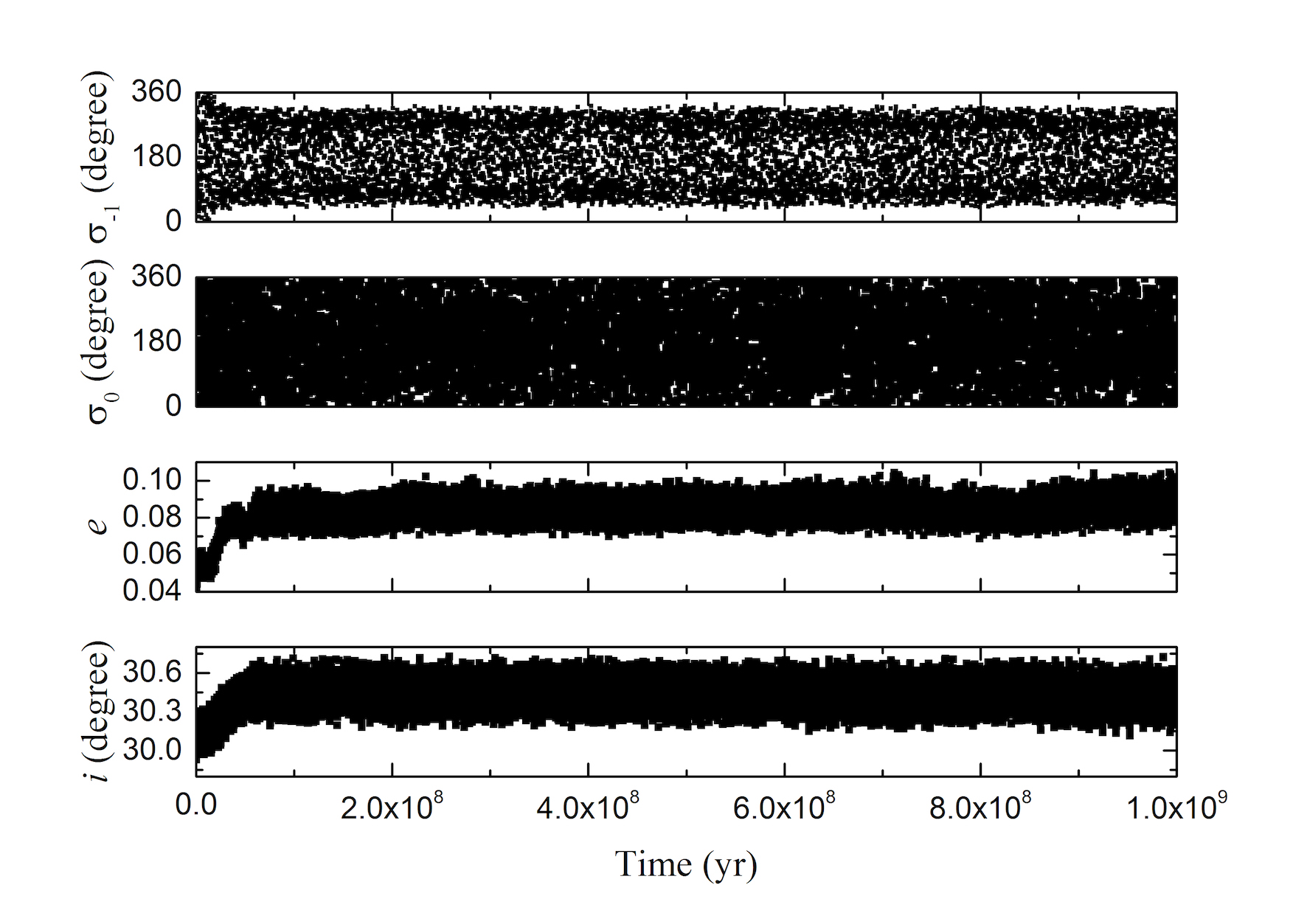}
  \caption{An example of the creation of a typical $-1$-mode 4:7 resonator starting with $E_0=0.05$ and $I_0=30^{\circ}$. This particle is indicated by one of the biggest green filled circles in Fig.~\ref{capture-1}, and unlike 2014 TZ85, is located inside the permissible region on the left side of the limiting curve. The four panels down from the top show the time evolution of the $-1$-mode 4:7 resonant angle $\sigma_{-1}$, the 0-mode 4:7 resonant angle $\sigma_0$ (which is circulating), the eccentricity $e$, and the inclination $i$. The resonant behavior is quite constant after Neptune's migration ceased at about $8\times10^7$ yr in the simulation.}
  \label{simulated-1}
\end{figure}

\subsubsection{Kozai mechanism}

Unlike the 0-mode 4:7 resonators, the $-1$-mode ones from both our simulations and the observation do not well follow the $(e, i)$ distribution predicted by the limiting curve $e_c^{-1}(i)$. As shown in Fig.~\ref{capture-1}, nearly half of the objects (including the real RKBO 2014 TZ85) are on the right side of the limiting curve, i.e., outside the permissible region for the libration of $\sigma_{-1}$. These odd $-1$-mode resonators with $e>e_c^{-1}$ can be explained by invoking the Kozai mechanism, as the same dynamical evolution scenario for 2014 TZ85 plotted in Fig.~\ref{2014TZ85}.  

In contrast, a representative example of typical $-1$-mode resonators with $e\le e_c^{-1}$ from our simulations is given in Fig.~\ref{simulated-1}. After capture into the 4:7 MMR at about $2\times10^7$ yr, the $-1$-mode resonant angle $\sigma_{-1}$ begins to librate stably until the end of the 1 Gyr integration, while the 0-mode resonant angle $\sigma_0$ is continuously circulating. The particle's $e$ and $i$ stay very constant after Neptune's migration has ceased at $\sim8\times10^7$ yr, with only very small amplitude cycling of $\Delta e<0.015$ and $\Delta i<0.4^{\circ}$. This orbital evolution indicates that the Kozai mechanism is not involved here, as such, the criterion $e\le e_c^{-1}$ has to be met for this stable $-1$-mode 4:7 resonator.  

We note that two of the $-1$-mode captured resonators are found to experience the $K_{0}$- and $K_{180}$-type Kozai mechanism, respectively, as indicated by $K_{0\&180}^{-1}$ in Table~\ref{resonators}.  Although the application of the limiting curve may fail for such $\omega$-librators (as discussed in Section 2.2), Fig.~\ref{simulated-1} shows that they both reside in the permissible region (denoted by the symbols $K_{0}^{-1}$ and $K_{180}^{-1}$). In addition, no $-1$-mode captured resonators associated with the $K_{90}$- or $K_{270}$-type Kozai mechanism are produced in our simulations.


\section{Conclusions and discussion}

Building on our previous works about the dynamics of Plutinos and Twotinos on inclined orbits \citep{Li2014a, Li2014b}, in this paper, we focused on the high-inclination population in the 4:7 MMR with Neptune. For this high-order resonance, there are two associated libration modes: (1) the ``0-mode'', which is the eccentricity-type resonance characterized by the libration of the resonant angle $\sigma_0$; (2) the  ``$-1$-mode'', which is the mixed-$(e, i)$-type resonance and only appears at high inclination. The $-1$-mode is defined by the independent libration of the resonant angle $\sigma_{-1}$, while $\sigma_0$ is simultaneously circulating.

\subsection{0-mode analysis: summary and predictions}

First, we studied the 0-mode of the 4:7 MMR semi-analytically. We measured the critical eccentricity $e_c^0$ as a function of the inclination $i$: if $e\ge e_c^0$, the libration of $\sigma_0$ is permitted. With increasing $i$, the value of $e_c^0$ will become larger, and this defines the limiting curve $e_c^0(i)$. We find that all the observed 0-mode 4:7 RKBOs indeed lie to the right side of this curve (i.e., satisfying $e\ge e_c^0(i)$), which is denoted as the permissible region (see Fig.~\ref{PR}(a)). Our results indicate that the more inclined 4:7 resonators tend to possess more eccentric orbits.

To test the semi-analytical prediction, we performed numerical simulations with the Sun and four Jovian planets, where the high-inclination particles near the 4:7 MMR were integrated for 10 Myr. We found that the $a$-ranges of the 0-mode resonant particles (i.e., the resonance boundaries) generally shrink with increasing initial inclinations $i_0$. For the lower initial eccentricities $e_0=0.1$ and 0.15, the 0-mode libration behavior disappears at $i_0>15^{\circ}$ and $>27^{\circ}$, respectively. For $e_0\ge0.2$, the resonance boundaries can extend to much higher inclinations, $i_0=30^{\circ}$. The numerical results are quantitatively consistent with our semi-analytical prediction that libration of the resonant angle $\sigma_0$ can only occur in the permissible zone, where $e_0\ge e_c^0(i_0)$.

For further investigation of the long-term evolution and orbital distribution of the 4:7 RKBOs, we conducted 4 Gyr simulations starting with the ($e_0, i_0$) pairs for particles with observed 0-mode libration in the shorter pre-runs. We found that although the resonant survivors could have quite large mobility in $(e, i)$ space due to chaotic evolution, at the end of the 4~Gyr simulations, nearly all of them still reside in the permissible region where the condition $e\ge e_c^0(i)$ is satisfied for the libration of $\sigma_0$. We also noticed that these resonant survivors spread out to cover nearly the entire permissible region, following the trend that the inclination increases with increasing eccentricity. The results indicate that the limiting curve $e_c^0(i)$ can be used to accurately predict the limits of the $(e, i)$ distribution of the real objects in the 0-mode 4:7 MMR. It must be mentioned that, in each case of $e_0=0.15$, 0.2, 0.25, there is a single simulated resonator out of hundreds that ended the integration outside the permissible region. These three particles leave the permissible region due to the libration of $\omega$ around $0^{\circ}$, i.e., the $K_{0}$-type Kozai mechanism, which was not involved in our semi-analytical calculation of $e_c^0$. Nevertheless, the probability of such exceptions is extremely low, $\lesssim1$\% in our simulations, and should also be rare for real 0-mode 4:7 RKBOs.

Next, we explored the scenario of the 4:7 MMR sweeping and capture during the final phase of Neptune's migration, for test particles initially distributed between 41.9 au and 44 au, with initial eccentricities $E_0=0.05-0.25$ and initial inclinations $I_0=1^{\circ}-30^{\circ}$. The results show that the captured resonators associated with the 0-mode 4:7 MMR are, without exception, found in the permissible region. Combining these results with the long-term evolution simulations, the particles swept-up during Neptune's migration can potentially survive in the resonance to the present day. We find that the particle population initially outside the permissible region, with $E_0<e_c^0(I_0)$, can be chaotically captured into the migrating 4:7 MMR at the very beginning, have $e$ excited inside the resonance, and achieve stable 0-mode resonant orbits after meeting the requirement of $e\ge e_c^0$. For high-inclination particles with $I_0\ge10^{\circ}$, the small $E_0$ appears to increase the probability of resonance capture, while large $E_0$ ($\ge e_c^0$) can help to increase the likelihood of remaining in the resonance. The competition between these two factors determines the number of the 0-mode 4:7 resonators at the end of the resonance sweeping process.

\begin{figure}
 \hspace{-0.7cm}
  \centering
  \includegraphics[width=9cm]{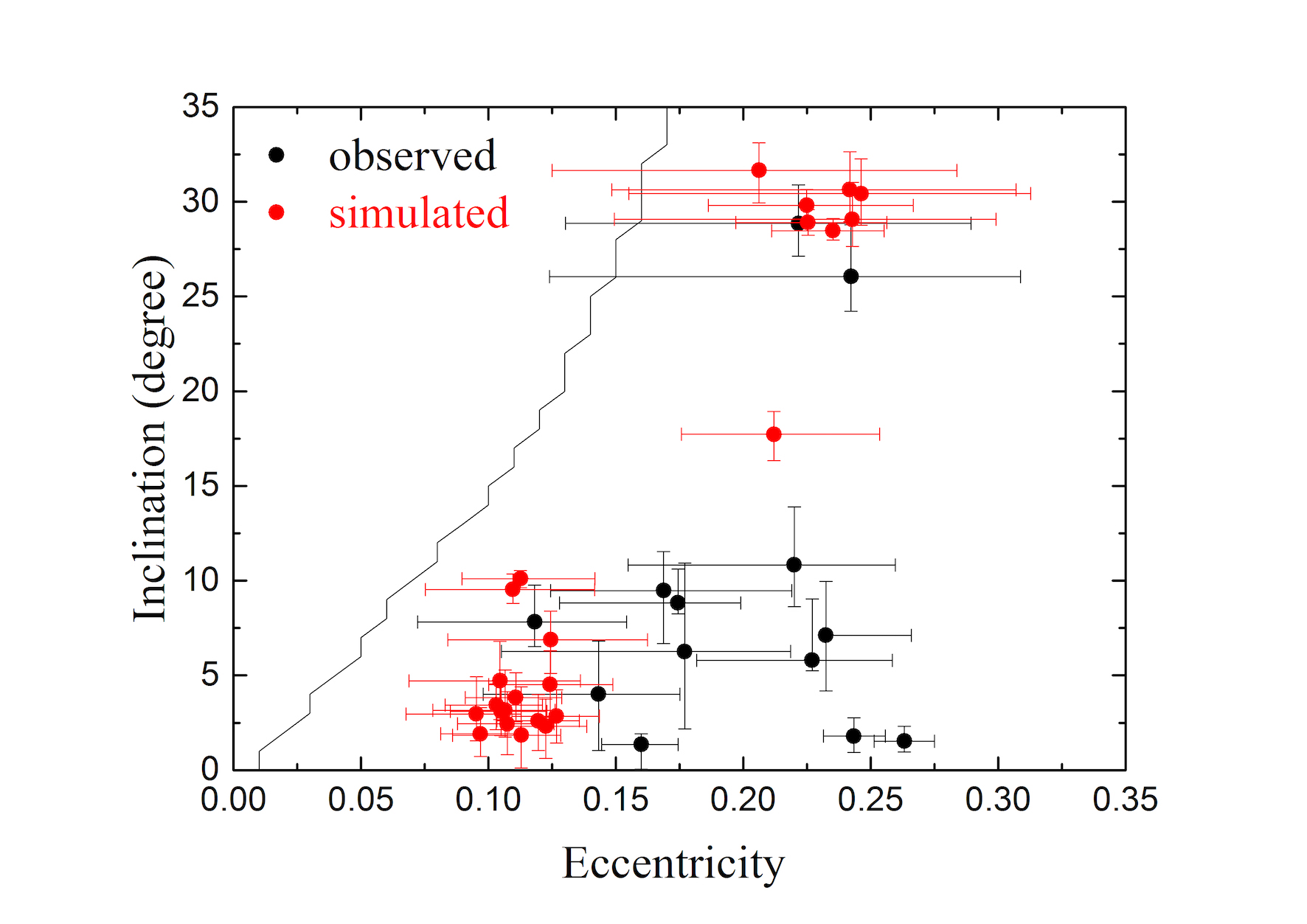}
  \caption{Distribution of eccentricities ($e$) and inclinations ($i$) for the Kozai objects associated with the 0-mode 4:7 MMR. These orbital elements are averaged over the 10 Myr integration, and error bars indicate the full amplitudes of the $e$- and $i$-oscillations. The black colour denotes the real observed samples from Fig.~\ref{PR}(a), and the red colour denotes the simulated samples generated in the planet migration model from Fig.~\ref{capture0}.}
  \label{Kerror}
\end{figure}

Among captured resonators in the 0-mode 4:7 MMR, there are a total of 8 objects having $\omega$ librating around $0^{\circ}$ or $180^{\circ}$. Such $\omega$-librators could possibly reside outside of the permissible region as we noted, but all our simulated particles in this state actually do not. This is likely because the particles with $e<e_c^0$ left the 4:7 MMR before the Kozai mechanism began operating. The other kind of Kozai libration, around $\omega=90^{\circ}$ or $270^{\circ}$, has 16 simulated resonators. We predict they are required to stay in the permissible region according to our resonant structure analysis, and indeed they do in our simulations. More interestingly, the $\omega$-librators can be observed to possess $i$ as small as $\lesssim2^{\circ}$. This suggests that, inside the 4:7 MMR, the Kozai mechanism may take place at any $i$-value; unlike in other lower lower-order resonances, where the Kozai mechanism can only operate at moderate-to-high $i$; or in the non-resonant Kuiper belt, where it is only observed at very high $i$.

We note that, for 4:7 resonators experiencing the Kozai mechanism, the typical oscillations of eccentricities and inclinations are $\sim0.01-0.12$ and $<5^{\circ}$, respectively. Fig.~\ref{Kerror} shows that a large fraction of the Kozai objects will always reside in the permissible region on the right side of the limiting curve during their oscillations. A few of the highest-$i$ ($\sim30^{\circ}$) particles may have considerable eccentricity oscillations, and the osculating $(e, i)$ can evolve outside the permissible region for a very short period of time. Such $e$- and $i$-oscillations can be seen for both the real, observed Kozai objects (black dots) and the simulated ones produced in the planet migration model (red dots). Here the use of averaged $(e, i)$, in order to be consistent with our semi-analytical calculations, has also the advantage of better constraining the distribution of the 4:7 RKBOs.

\subsection{$-1$-mode analysis: summary and predictions}

We then studied the $-1$-mode of the 4:7 MMR, though there is only a single KBO (2014 TZ85) known to be librating in this mode to date. In the planet migration model, we find that test particles starting at $I_0\ge10^{\circ}$ can be possibly captured into this mode. Among them, those with more inclined orbits of $i\gtrsim20^{\circ}$ could remain relatively stable up to 1 Gyr, like the real object 2014 TZ85 ($i\sim18^{\circ}$). In our simulations, the number ratio of captured resonators in the $-1$- and 0-mode is $N_{4:7}^{-1}$/$N_{4:7}^{0}\sim0.17$, which is high compared to observations: presently only 1 out of the 34 known 4:7 RKBOs is associated with the $-1$-mode. Finally, we note that because of the Kozai mechanism, the $(e, i)$ distribution of the simulated and real $-1$-mode resonators do not nicely abide by our predicted limiting curve $e_c^{-1}(i)$.

\subsection{Future observations}

Future observations specifically targeting high-$i$ RKBOs would help in measuring a more accurate value of the intrinsic number ratio $N_{4:7}^{-1}$/$N_{4:7}^{0}$, which is currently observationally biased toward low-$i$ RKBOs due to most KBO surveys being constrained to the ecliptic plane. Further studies with enough 4:7 RKBOs to statistically measure this ratio may provide a new prediction which future observations can use to constrain Neptune's migration and the orbital distribution of the primordial planetesimal disk.


\section*{Acknowledgments}

This work was supported by the National Natural Science Foundation of China (NSFC, Nos. 11973027, 11473015, 11473016 and 11933001), and the Fundamental Research Funds for the Central Universities (No. 020114380024). The authors would like to express their sincere thanks to an anonymous referee for the critical comments that helped to considerably improve the manuscript, and also to another referee, Prof. Valerio Carruba, for his careful reading of this paper as well as valuable comments.

\end{document}